\documentclass[a4paper,fleqn,usenatbib]{mnras}

\usepackage{newtxtext,newtxmath}

\usepackage[T1]{fontenc}
\usepackage{ae,aecompl}

\usepackage{graphicx}	
\usepackage{amsmath}	
\usepackage{amssymb}	

\title[Rotational disruption and grain size distribution]{Effects of rotational disruption on the
evolution of grain size distribution in galaxies}

\author[H. Hirashita and T. Hoang]{
Hiroyuki Hirashita$^1$\thanks{E-mail: hirashita@asiaa.sinica.edu.tw} and
Thiem Hoang$^{2,3}$
\\
$^{1}$Institute of Astronomy and Astrophysics, Academia Sinica,
Astronomy-Mathematics Building, AS/NTU,\\
No.\ 1, Sec.\ 4, Roosevelt Road, Taipei 10617, Taiwan \\
$^{2}$Korea Astronomy and Space Science Institute, Daejeon 34055, Republic of Korea\\
$^3$Korea University of Science and Technology, 217 Gajeong-ro, Yuseong-gu, Daejeon 34113,
Republic of Korea}

\date{Accepted XXX. Received YYY; in original form ZZZ}

\pubyear{2020}

\begin{document}
\label{firstpage}
\pagerange{\pageref{firstpage}--\pageref{lastpage}}
\maketitle

\begin{abstract}
Interstellar dust grains can be spun up by radiative torques, and
the resulting centrifugal force may be strong enough to disrupt large dust grains.
We examine the effect of this rotational disruption on the evolution of grain size
distribution in galaxies. To this goal, we modify our previous model
by assuming that rotational disruption is the major small-grain production mechanism.
We find that rotational disruption can have a large influence on the evolution of
grain size distribution in the following two aspects especially for composites and
grain mantles (with tensile strength $\sim 10^7$ erg cm$^{-3}$).
First, because of the short time-scale of rotational disruption, the small-grain
production occurs even in the early phase of galaxy evolution.
Therefore, even though stars produce large grains, the abundance of small grains
can be large enough to steepen the extinction curve.
Secondly, rotational disruption is important in determining the maximum grain radius,
which regulates the steepness of the extinction curve.
For compact grains with tensile strength $\gtrsim 10^9$ erg cm$^{-3}$,
the size evolution is significantly affected by rotational
disruption only if the radiation field is as strong as (or the dust temperature is
as high as) expected for starburst galaxies. For compact grains, rotational
disruption predicts that
the maximum grain radius becomes less than 0.2 $\micron$ for galaxies with
a dust temperature $\gtrsim 50$ K.
\end{abstract}


\section{Introduction}\label{sec:intro}

Dust grains are known to be important for various processes in the
interstellar medium (ISM) and galaxies.
Many of the important processes concern the surface area of dust grains.
Some molecules such as H$_2$ form on dust surfaces
\citep[e.g.][]{Gould:1963aa,Cazaux:2004aa}; thus, their
reaction rates are governed by the grain surface area, which is determined by
the grain size distribution (distribution function of grain radius)
\citep{Hirashita:2017aa,Chen:2018aa}.
Dust grains also
interact with radiation by absorbing stellar ultraviolet (UV)--optical light and
reprocess it in the infrared. This means that dust modifies or even
governs the
spectral energy distributions (SEDs) of galaxies
\citep[e.g.][]{Takeuchi:2005aa}.
The cross-section of a dust grain for absorption and scattering (i.e.\ extinction),
whose wavelength dependence is referred to as the extinction curve,
depends on the grain size distribution \citep[e.g.][hereafter MRN]{Mathis:1977aa}.
Thus, the evolution of grain size distribution is of fundamental importance in understanding
the chemical and radiative processes in the ISM and galaxies.

There have been some studies modelling and calculating the evolution of
grain size distribution in the ISM.
\citet{Liffman:1989aa} traced
the size evolution of individual dust grains, considering dust destruction by
supernova (SN) shocks and dust growth by accretion. This method is useful, for example, to
derive the lifetime of dust as a function of grain radius, but the interactions between grains
are not easy to include.
\citet{ODonnell:1997aa} modelled the grain size distribution by further considering
collisional processes (coagulation and
shattering) in a multiphase ISM and showed that the extinction curves calculated from
the resulting grain size distributions broadly reproduce the Milky Way extinction curve.
The evolution of grain size distribution was first modelled in a consistent manner with the
chemical enrichment of galaxies by \citet{Yamasawa:2011aa}.
They took into account dust formation and destruction by SNe.
In addition to these processes, \citet{Asano:2013aa} completed the model by
additionally including
dust formation by AGB stars, dust growth by accretion and coagulation, and dust disruption
by shattering.
\citet{Nozawa:2015aa} applied this model to high-redshift quasars, and showed that the
calculated grain size distribution explains the observed extinction curve in a
$z=6.2$ quasar ($z$ is the redshift).
The evolution model of grain size distribution has been further combined with
hydrodynamic simulations by post-processing \citep{Hirashita:2019aa,Rau:2019aa}
or by a direct implementation \citep{McKinnon:2018aa,Aoyama:2020aa}.
\citet[][hereafter HM20]{Hirashita:2020aa} used a one-zone galaxy model but
further decomposed the calculated grain size distribution into relevant
species (silicate and aromatic/non-aromatic carbon).
Their model is also consistent with the abundance of polycyclic aromatic hydrocarbons (PAHs)
in nearby galaxies \citep[see also][]{Seok:2014aa}.

In the above, shattering is the unique mechanism that creates small grains efficiently.
Recently, \citet{Hoang:2019aa} has proposed another mechanism of creating small grains
from large grains -- rotational disruption \citep[see also][]{Hoang:2019ab}.
In the presence of strong radiation, grains acquire suprathermal rotation driven by radiative torques
\citep{Draine:1996ab,Hoang:2008aa,Hoang:2009aa}. This rotation induces the centrifugal stress in the
dust grain.
\citet{Hoang:2019ab} found that large dust grains can be disrupted because the centrifugal stress
exceeds the tensile strength.
\citet{Hoang:2019aa} further showed that rotational disruption determines the maximum grain radius
(or the upper cut-off of the grain size distribution),
which depends on both the interstellar radiation field (ISRF) intensity and the grain tensile strength.
Other than determining the upper cut-off of grain radii, rotational disruption could also enhance
the small grain production.
Thus, for a complete understanding of the grain size distributions in galaxies, we should clarify
how rotational disruption affects the evolution of grain size.

Since rotational disruption is more efficient in higher ISRFs, it is more important in starburst
(or actively star-forming) galaxies. The dust temperature which
reflects the ISRF intensity is generally high in those galaxies \citep[e.g.][]{Zavala:2018aa}.
Starburst galaxies emit most of their stellar radiation energy by dust emission
\citep{Sanders:1996aa}, which
means that understanding dust properties is crucial in clarifying the radiative
processes in those galaxies.
In addition, the fraction of stellar light reprocessed by dust increases towards $z\sim 1$--2,
when the cosmic star formation rate is the highest
\citep[e.g.][]{Takeuchi:2005ab,Goto:2010aa,Burgarella:2013aa}.
In such dusty star-forming galaxies, rotational disruption could play a significant role in
determining the dust properties through the modification of grain size distribution.

The goal of this paper is to clarify the effect of rotational disruption on
the evolution of grain size distribution. {As a representative quantity that
reflects the grain size distribution, we also calculate extinction curves.}
In our previous modelling, shattering was the unique process that produces small grains, but
the shattering
rate might be uncertain because of the assumed grain velocities.
In this paper, we treat rotational disruption as \textit{an alternative mechanism of small-grain production}
(i.e.\ we examine the case where rotational disruption is the dominant mechanism of creating
small grains from
large grains). This step also serves to provide a means of testing the role of rotational
disruption against the observed dust properties.

To this goal, we newly
formulate and include the effect of rotational disruption in the framework that calculates
the evolution of grain size distribution in a galaxy. The development of the framework is
based on HM20, in which we
also separated the grains into relevant species to calculate the extinction curve. This
enables us to
examine the effect of rotational disruption on the extinction curve as an observable
signature. We also discuss the effects of a strong ISRF as expected for starburst galaxies and
some high-redshift galaxies.

This paper is organized as follows.
In Section~\ref{sec:model}, we describe the dust evolution model which treats the
grain size distribution and the grain composition. In particular, we newly include the effect of
rotational disruption on the grain size distribution.
In Section \ref{sec:result}, we show the results.
In Section \ref{sec:discussion}, we provide some extended discussions, especially, regarding
the significance of rotational disruption in galaxies with high ISRF intensities.
In Section \ref{sec:conclusion}, we give the conclusion of this paper.

\section{Model}\label{sec:model}

{The main purpose of the modeling in this paper is to investigate rotational
disruption as an alternative mechanism of shattering in producing small grains.
For this purpose, we use and extend the framework developed in our previous papers
(HM20, originally based on \citealt{Hirashita:2019aa}).}
We model the metal and dust enrichment in a galaxy, which is treated as a one-zone
object. The evolution of grain size distribution is calculated in a manner consistent with
dust enrichment and interstellar dust processing.
We newly include the effect of rotational disruption.
Shattering also has a similar effect on small-grain production; in this paper, in order
to clarify the role of rotational disruption in small-grain production,
we turn off shattering.
In other words, we include rotational disruption as an alternative small-grain-production
mechanism to shattering to examine if rotational disruption could play a significant role in
reproducing the observed dust properties.
This is also equivalent to the assumption that rotational disruption
is the dominant small-grain production mechanism over shattering.
Turning off shattering also serves to obtain results not affected by the uncertainty in
shattering (mainly caused by the assumed turbulence model).
The case where rotational disruption and shattering coexist is discussed separately in
Section \ref{subsec:shattering}.

The grain size distribution is further decomposed into silicate and carbonaceous dust;
the latter species is further separated into aromatic and non-aromatic components.
Distinction among the grain species is based on HM20 and is necessary to calculate
the extinction curve.
The infrared SED of dust emission could also be a useful output; however, since
the calculation requires
a consistent treatment of extinction curves and radiation transfer effects,
the infrared SED is left for a future work.
We only describe the outline of the models, and refer the interested reader to
HM20 and \citet{Hirashita:2019aa} for details.

\subsection{Evolution of grain size distribution}\label{subsec:size}

We assume
the galaxy to be a closed box; that is, the galaxy starts its evolution
with a certain mass of gas with zero metallicity and converts the gas to stars with
the total baryonic (gas + stars) mass conserved.
We adopt the following simple functional form for the star formation rate,
$\psi$:\footnote{In HM20, there is a typo for this equation.}
\begin{align}
\psi (t)=\frac{M_\mathrm{g,0}}{\tau_\mathrm{SF}}\exp\left( -\frac{t}{\tau_\mathrm{SF}}\right),
\end{align}
where $M_\mathrm{g,0}$ is the initial gas mass (= total baryonic mass) and
$\tau_\mathrm{SF}$ is the star formation time-scale given as a free parameter.
Using a set of equations that describe the chemical evolution of the galaxy,
we obtain the time evolution of the masses of gas, stars, and metals,
denoted as $M_\mathrm{gas}$, $M_\star$, and $M_Z$, respectively.
Because of the above assumption of closed box, $M_\mathrm{gas}+M_\star =M_\mathrm{g,0}$
($M_\star =0$ at $t=0$).
The most important quantity that is calculated by this chemical evolution framework is
the metallicity $Z\equiv M_Z/M_\mathrm{gas}$. We also compute the silicon and
carbon abundances (relative to the total gas mass), which are denoted as
$Z_\mathrm{Si}$ and $Z_\mathrm{C}$, respectively. In calculating the abundances of
metals, silicon and carbon, we adopt the stellar yields in the literature
(see HM20 for the references). We also need the SN rate (denoted as $\gamma$), which is calculated by
assuming that stars in a zero-age-main-sequence mass range of 8--40 M$_{\sun}$ become
SNe at the end of their lives.
We adopt the Chabrier initial mass function \citep{Chabrier:2003aa} 
with a stellar mass range of 0.1--100 M$_{\sun}$.

We calculate the evolution of grain size distribution by considering the following processes:
stellar dust production, dust destruction by SN shocks in
the ISM, dust growth by accretion and grain--grain sticking by coagulation.
As mentioned above, we turn off shattering to clarify the importance of small grain production
by rotational disruption.
We assume grains to be spherical and compact,
so that the grain mass $m$ is related to the grain radius $a$ as $m=(4\upi /3)a^3s$,
where $s$ is the material density of dust.
We adopt $s=2.24$ g cm$^{-3}$, which is appropriate for graphite
\citep{Weingartner:2001aa}.
As explained in HM20, our model is not able to
include multiple dust components directly in the calculation of grain size distribution
because of the difficulty in treating the interactions among various grain species.
Thus, we adopt a
`representative' species for the grain size distribution, which is later separated into
various species when we calculate the extinction curve. HM20 confirmed
that the results are not sensitive to the adopted representative species.
The grain size distribution at time $t$ is expressed by the grain mass distribution
$\rho_\mathrm{d}(m,\, t)$, which is defined such that
$\rho_\mathrm{d}(m,\, t)\,\mathrm{d}m$ is the mass density
of dust grains whose mass is between $m$ and $m+\mathrm{d}m$.
The grain size distribution,
$n(a,\, t)$, is derived from the grain mass distribution as
\begin{align}
\rho_\mathrm{d}(m,\, t)\,\mathrm{d}m=\frac{4}{3}\upi a^3sn(a,\, t)\,\mathrm{d}a.\label{eq:rho_vs_n}
\end{align}

For dust processing, a multi-phase nature of the ISM is necessary to model.
We assume that the ISM is composed of the diffuse (warm) and dense (cold) components,
which have
$(n_\mathrm{H}/\mathrm{cm}^{-3},\, T_\mathrm{gas}/\mathrm{K})=(0.3,\, 10^4)$
and $(300,\, 25)$, respectively
($n_\mathrm{H}$ is the hydrogen number density and $T_\mathrm{gas}$ is the gas temperature).
The mass fraction of the dense ISM is given as a constant parameter,
$\eta_\mathrm{dense}$.
We calculate the change of grain mass distribution in a time-step interval $\Delta t$ as
$\Delta{\rho}_\mathrm{d}(m,\, t)=[{\upartial{\rho}_\mathrm{d}(m,\, t)}/{\upartial t}]_i\,f_i\Delta t$,
where $i$ indicates each process,
$[{\upartial{\rho}_\mathrm{d}(m,\, t)}/{\upartial t}]_i$ is the contribution from
process $i$ to the change of the grain mass distribution per unit time,
and $f_i$ is the fraction of the gas phase that hosts the process.
The processes we consider are stellar dust production ($i=\mbox{star}$), SN destruction by sputtering
($i=\mbox{sput}$),
dust growth by accretion ($i=\mbox{acc}$),
and grain growth by coagulation ($i=\mbox{coag}$). We newly include
rotational disruption ($i=\mbox{disr}$) in this paper.
Stellar dust production and SN destruction are assumed to occur in both
ISM phases, so that $f_\mathrm{star}=f_\mathrm{sput}=1$.
Coagulation and accretion take place only in the dense phase,
so that $f_\mathrm{coag}=f_\mathrm{acc}=\eta_\mathrm{dense}$.
We also assume that rotational disruption occurs only in the
diffuse ISM ($f_\mathrm{disr}=1-\eta_\mathrm{dense}$) for the following two reasons.
First, the damping of grain rotation by sticking collision with gas particles
is significant in the
dense ISM \citep{Hoang:2019aa}. Secondly, the radiation could be effectively attenuated in the
dense ISM. However, since rotational disruption takes place on a much shorter time-scale than
the dust enrichment processes, the choice of $f_\mathrm{disr}$ does not affect our conclusions.

For the stellar dust production, we adopt the dust yields of SNe and AGB stars in the
literature (see HM20 for the references). The dust mass supplied by the stellar sources are
distributed in each grain radius bin based on the lognormal grain size distribution centered
at 0.1 $\micron$ with standard deviation $\sigma =0.47$ \citep{Asano:2013aa}.
Thus, in our model, we assume that stars produce large ($a\sim 0.1~\micron$) grains
(see HM20 for further arguments supporting this assumption).

The changes of grain size distribution by SN dust destruction and dust growth (accretion)
are treated by taking the grain-size-dependent (or grain-mass-dependent) destruction and
growth time-scales into account. The evolution of grain size distribution by these processes
is described by an `advection' equation in the grain-radius space
\citep{Hirashita:2019aa}.
The change of grain size distribution by coagulation is, on the other hand,
described by the Smoluchowski equation.
The grain--grain collision rates for various combinations of grain radii are
evaluated based on the geometric cross-section and the turbulence-driven relative grain velocity
\citep{Ormel:2009aa}.
The sticking efficiency is assume to be unity.
The evolution of grain size distribution by rotational disruption is also calculated as
described in the next subsection.


\subsection{Effect of rotational disruption}\label{subsec:rot}

Rotational disruption causes fragmentation of grains by centrifugal force.
The mechanism of disruption is common between rotational disruption and shattering
in the sense that the internal stress exceeding the tensile
strength is the cause of the disruption.
Therefore, we expect that the grain size distribution is modified by rotational disruption
in a way similar to shattering. Rotational disruption does not require grain--grain
collisions, but occurs spontaneously on a time-scale (as a function of grain mass)
$\tau_\mathrm{disr}(m)$, which
is evaluated later. The evolution of grain mass distribution by rotational disruption is
described as
\begin{align}
\left[\frac{\upartial\rho_\mathrm{d}(m,\, t)}{\upartial t}\right]_\mathrm{disr}
= -\frac{\rho_\mathrm{d}(m,\, t)}{\tau_\mathrm{disr}(m)}
+ \int_0^\infty\frac{\rho_\mathrm{d}(m_1,\, t)}{m_1\tau_\mathrm{disr}(m_1)}\,
\theta_\mathrm{frag}(m;\, m_1)\,\mathrm{d}m_1
,\label{eq:dis}
\end{align}
where $\theta_\mathrm{frag}(m;\, m_1)$ is the mass distribution function of fragments
produced from a grain with mass $m_1$. Although this equation can be derived for its own sake,
it may be convenient to describe it in correspondense with the equation governing shattering
as described in Appendix.

We adopt the following time-scale for rotational disruption from equation (32) of
\citet{Hoang:2019aa}:
\begin{align}
\tau_\mathrm{disr}(a) &= 10^5U^{-1}\left(\frac{S_\mathrm{max}}{10^7~\mathrm{erg~cm}^{-3}}\right)^{1/2}
\left(\frac{a}{0.1~\micron}\right)^{-0.7}~\mathrm{yr}\nonumber\\
& \mbox{~~for $a\geq a_\mathrm{disr}$},\label{eq:tau_disr}
\end{align}
where $U$ is the ISRF intensity normalized to the average value in the
solar neighbourhood as given by \citet{Mathis:1983aa}, $S_\mathrm{max}$ is
the tensile strength, and $a_\mathrm{disr}$ is the threshold grain radius for rotational disruption.
We fixed the mean wavelength of the ISRF to $\bar{\lambda}=0.5~\micron$ in the
original expression. For $a<a_\mathrm{disr}$, no rotational disruption is assumed to occur
(i.e.\ $\tau_\mathrm{disr}=\infty$).

The threshold grain radius, $a_\mathrm{disr}$, is estimated by \citet{Hoang:2019aa}.
We can use their equation (29), since the gas density and temperature of the diffuse ISM
are in the regime where the damping of the grain
rotation by the emission of IR photons is important. We also fix the radiation anisotropy
degree to $\gamma_\mathrm{rad}=0.1$ and the mean ISRF wavelength to $\bar{\lambda}=0.5~\micron$,
obtaining
\begin{align}
\left(\frac{a_\mathrm{disr}}{0.1~\micron}\right)^{2.7}=5.1U^{-1/3}\left(
\frac{S_\mathrm{max}}{10^7~\mathrm{erg~cm}^{-3}}\right)^{1/2}.\label{eq:a_disr}
\end{align}
Note that this is valid for $a_\mathrm{disr}\lesssim \bar{\lambda}/1.8\simeq 0.28~\micron$
(i.e.\ in the regime where the efficiency of radiative torque strongly depends on the grain radius).
If $a_\mathrm{disr}$ is larger than 0.3 $\micron$, radiative torque cannot make the grain
rotation fast enough to cause rotational disruption \citep{Hoang:2019aa}.

Finally, we determine the fragment mass distribution, $\theta_\mathrm{frag}(m;\, m_1)$.
{Since there is no available physical model for the fragment mass distribution
in rotational disruption,
we use the fragment distribution function in shattering as a guide.}
The maximum and minimum masses of the fragments produced from a grain mass
$m_1$ are assumed to be
$m_\mathrm{f,max}=0.02m_1$ and
$m_\mathrm{f,min}=10^{-6}m_\mathrm{f,max}$, respectively
\citep{Guillet:2011aa,Hirashita:2019aa}.
We adopt the following mass distribution function of fragments produced from
a grain with mass $m_1$:
\begin{align}
\theta_\mathrm{frag}(m;\, m_1) &=
\frac{(4-\alpha_\mathrm{f})m_1m^{(-\alpha_\mathrm{f}+1)/3}}{3\left[
m_\mathrm{f,max}^\frac{4-\alpha_\mathrm{f}}{3}-
m_\mathrm{f,min}^\frac{4-\alpha_\mathrm{f}}{3}\right]}\,
\Phi (m;\, m_\mathrm{f,min},\, m_\mathrm{f,max}),\label{eq:frag}
\end{align}
where
$\Phi (m;\, m_\mathrm{f,min},\, m_\mathrm{f,max})=1$ if
$m_\mathrm{f,min}\leq m\leq m_\mathrm{f,max}$, and 0 otherwise,
and
$\alpha_\mathrm{f}=3.3$ \citep{Jones:1996aa}.

{The above fragment mass distribution could be optimistic in the number of small fragments,
but we basically adopt it to investigate a possibility of rotational disruption being
an alternative to shattering for small-grain production. Formation of many small fragments
in rotational disruption may be justified for inhomogeneous grains
such as composite grains composed of tiny monomers and grains covered by mantles.
Such inhomogeneous structures generally contain many `interfaces' with locally weak binding force.
Since we are particularly interested in the disruption of highly inhomogeneous grains
(with small tensile strength as explained in Section \ref{subsec:param}), it would be appropriate to
consider the formation of many small fragments.
We should also keep in mind that
in the above, the whole disrupted grain is assumed to be fragmented
without leaving a remnant. Thus, our model provides an optimistic case for the
production of small grains also in this sense.
We discuss the dependence on the fragment size distribution later
in Section \ref{subsec:robustness}.}

\subsection{Calculation of extinction curve}\label{subsec:ext_method}

As an observable quantity, we calculate the extinction curve.
For this purpose, we decompose
the grain size distribution into the relevant dust species.
We adopt the same dust species as in HA20 for the
convenience of comparison.
The grain mass distributions of silicate ($\rho_\mathrm{sil}$) and
carbonaceous dust ($\rho_\mathrm{car}$) are
described by
$\rho_\mathrm{sil}(m,\, t) = f_\mathrm{sil}(t)\rho_\mathrm{d}(m,\, t)$, and
$\rho_\mathrm{car}(m,\, t) = [1-f_\mathrm{sil}(t)]\rho_\mathrm{d}(m,\, t)$, respectively,
where $f_\mathrm{sil}(t)$ is the mass ratio of silicate to the total dust mass.
The silicate fraction is calculated using the silicon-to-carbon abundance ratio
calculated in the chemical evolution model in Section~\ref{subsec:size}.

The carbonaceous grain size distribution is separated into
aromatic and non-aromatic populations.
We denote the grain mass distributions of the aromatic and non-aromatic species as
$\rho_\mathrm{ar}(m,\, t)$ and and $\rho_\mathrm{non-ar}(m,\, t)$, respectively,
and introduce the aromatic fraction, which is defined as
$f_\mathrm{ar}(m,\, t)\equiv\rho_\mathrm{ar}(m,\, t)/\rho_\mathrm{car}(m,\, t)$.
HA20 solved aromatization and aliphatization, and
showed that the aromatic fraction is $\simeq 1-\eta_\mathrm{dense}$ for most of the
grain size range of interest. Therefore, we simply assume a constant aromatic fraction,
$f_\mathrm{ar}=1-\eta_\mathrm{dense}$; that is,
$\rho_\mathrm{ar}(m,\, t)=(1-\eta_\mathrm{dense})\rho_\mathrm{car}(m,\, t)$ and
$\rho_\mathrm{non-ar}(m,\, t)=\eta_\mathrm{dense}\rho_\mathrm{car}(m,\, t)$.

The extinction at wavelength $\lambda $ in units of magnitude ($A_{\lambda }$)
is calculated as
\begin{align} 
A_{\lambda}=(2.5\log_{10} \mathrm{e})L\displaystyle\sum_{j}\displaystyle\int_{0}^{\infty}
n_{j}(a)\,\upi a^{2}Q_{\rm ext}(a, \lambda),
\end{align}
where the subscript $j$ indicates the grain species (i.e.\
silicate, aromatic carbon, and non-aromatic carbon), $L$ is the path length,
and $Q_{\rm ext}(a, \lambda)$ is the extinction efficiency factor evaluated
by using the Mie theory \citep[][]{Bohren:1983aa}.
We use astronomical silicate \citep{Weingartner:2001aa} for silicate, while
we adopt graphite in the same paper for
aromatic carbonaceous grains.
For non-aromatic species,
we adopt the optical constants of amorphous carbon
taken from \citet{Zubko:1996aa} (their ACAR)
\citep[see also][]{Nozawa:2015aa,Hou:2016aa}.
HM20 also provided extinction curves calculated with different carbonaceous
species; we only examine the above set of materials since it was successful
in producing extinction curves similar to the Milky Way curve (in particular,
the UV slope and the 2175 \AA\ bump).
The grain size distribution can be calculated by
$n_j(a,\, t)=3\rho_j(m,\, t)/a$ (see equation \ref{eq:rho_vs_n}).
The extinction is normalized to the value in the $V$ band
($\lambda ^{-1}=1.8\,\mu {\rm m}^{-1}$); that is, we output $A_\lambda /A_V$,
where $L$ is cancelled out.


\subsection{Variation of parameters}\label{subsec:param}

In this paper, we concentrate on the parameters that directly affect rotational disruption,
i.e.\ $U$ and $S_\mathrm{max}$.
We fix the dense gas fraction ($\eta_\mathrm{dense}=0.5$)
and the star formation time-scale ($\tau_\mathrm{SF}=5$ Gyr) unless otherwise stated.
These values are appropriate for Milky Way-like galaxies
\citep{Nozawa:2015aa}.

{The ISRF intensity $U$ strongly depends on the star formation activity;
however, it is also affected by the spatial dust distribution and the shielding
of dust (i.e.\ by various radiation transfer effects).}
Since our one-zone model is not suitable for modeling radiation transfer,
we simply treat $U$ as a free parameter.
{We also neglect the spatial inhomogeneity in $U$ and concentrate on the mean ISRF in the
galaxy. Since rotational disruption occurs on a short time-scale, the local ISRF could be important.
If the ISRF has a large variation within the galaxy, rotational disruption acts with different magnitudes
among the places. Our one-zone treatment is not able to address this issue; thus,
any effects that require spatially resolved treatments (such as the spatial and temporal inhomogeneity in
the radiation field that dust grains experience)
are left for a future work. This issue could be investigated by using radiation-hydrodynamic simulations that
consistently solve the dust evolution \citep{McKinnon:2020aa}.
We emphasize that our simple treatment for $U$ enables us to draw an explicit conclusion regarding
the dependence on
$U$, which could be useful for interpreting galaxies with known $U$ (note that $U$ can be
estimated from the dust temperature; see Section \ref{subsec:param_space}).}
We mainly examine the range of
$U\sim 1$--100, which is appropriate for
nearby galaxies \citep[e.g.][]{Hirashita:2009ac},
except for the starburst model discussed later.

For the tensile strength, the uncertainty is large because we only have limited
knowledge on the grain structures. Note that the grains larger than
$a_\mathrm{disr}$ are relevant here, since only grains with $a>a_\mathrm{disr}$ are
disrupted by rotational disruption. Therefore, to specify $S_\mathrm{max}$, we
have to consider the tensile strength of large grains.
Large grains could have complicated structures especially if they are formed
after accretion and coagulation.
Composite grains (composed of tiny monomers)
typically have $S_\mathrm{max}\sim 10^7$ erg cm$^{-3}$, while compact grains
have $S_\mathrm{max}\sim 10^9$ erg cm$^{-3}$--10$^{11}$ erg cm$^{-3}$
\citep{Hoang:2019ab,Hoang:2019aa}.
A core-mantle interface is expected to have a tensile strength similar to
composite grains.
Since the grain properties are unknown, we simply survey a wide range of
parameter for $S_\mathrm{max}\geq 10^7$ erg cm$^{-3}$.

\section{Results}\label{sec:result}

\subsection{Grain size distribution}\label{subsec:size_result}

\begin{figure}
\includegraphics[width=0.45\textwidth]{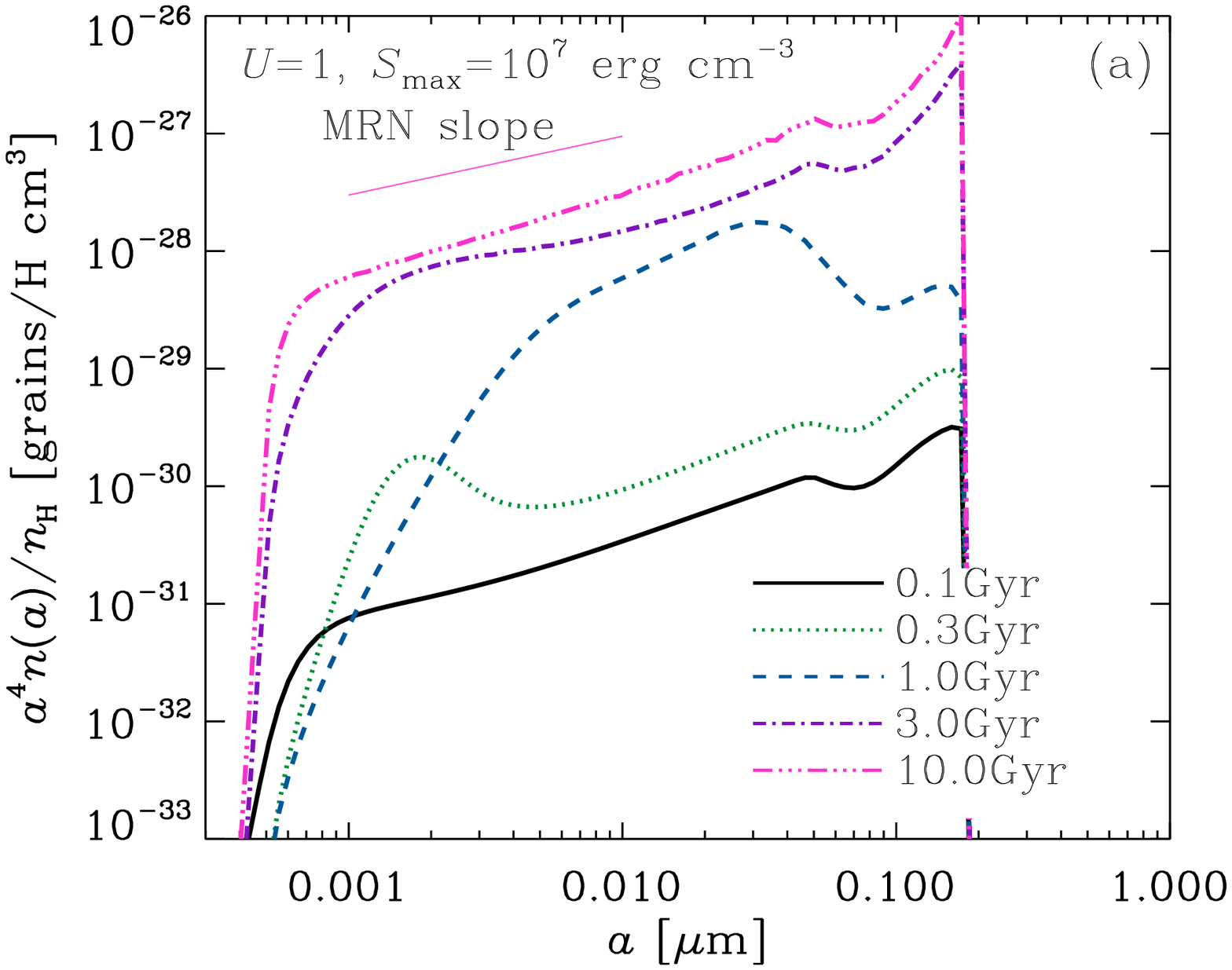}
\includegraphics[width=0.45\textwidth]{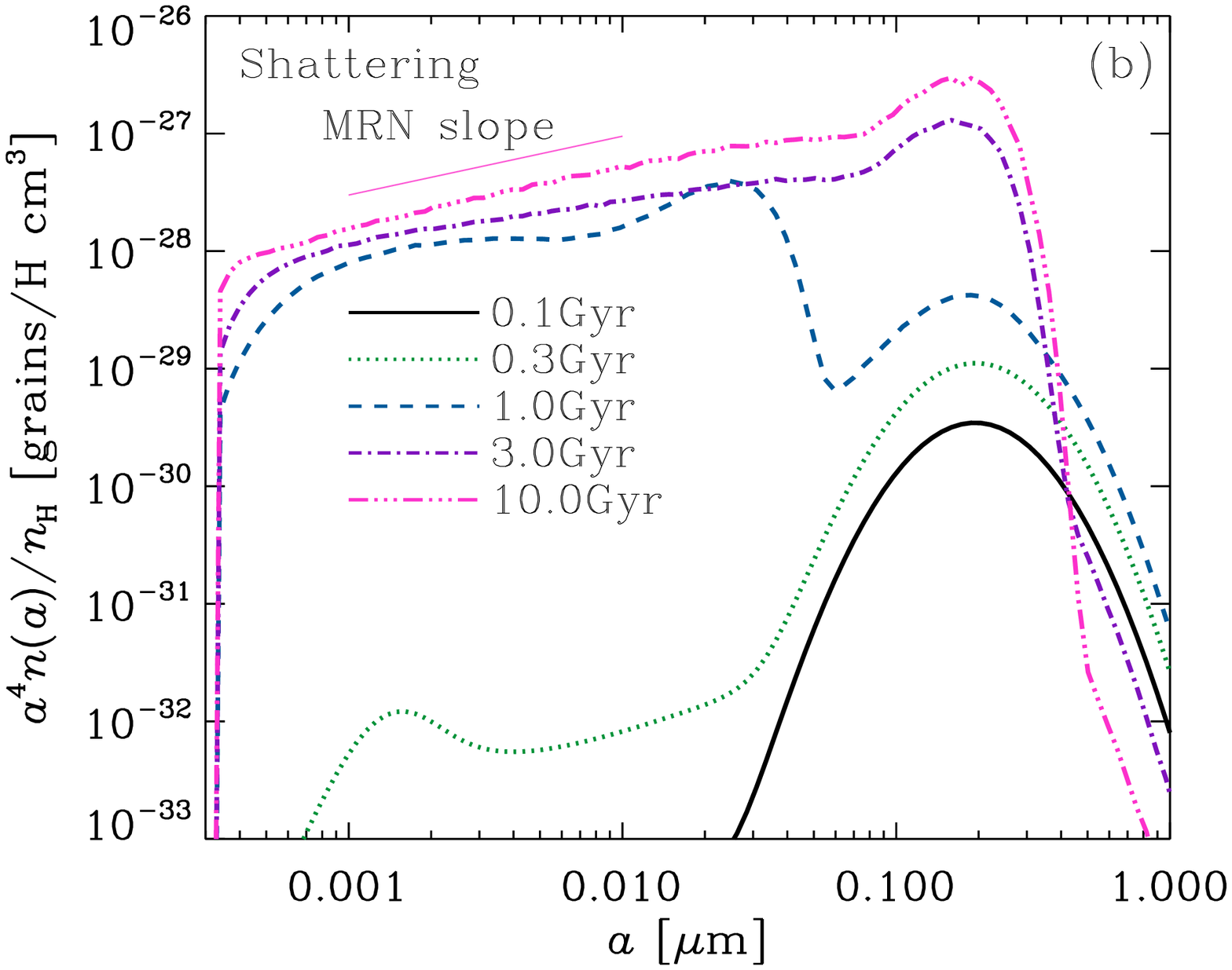}
\caption{Evolution of grain size distribution.
The grain size distribution is multiplied by $a^4$ and divided by $n_\mathrm{H}$,
so that the resulting quantity is proportional to the grain abundance  per $\log a$ relative to the
gas mass.
We show the results (a) with rotational disruption
and (b) with shattering instead of rotational disruption.
The solid, dotted, dashed, dot--dashed, and triple-dot--dashed lines show the results
at $t=0.1$, 0.3, 1, 3, and 10~Gyr, respectively.
The thin dotted straight line shows the slope of the MRN
grain size distribution ($n\propto a^{-3.5}$).
\label{fig:size}}
\end{figure}

We show the evolution of grain size distribution. We adopt $U=1$ for the `fiducial' case.
To investigate the case where the effect of rotational disruption
is the most prominently seen, we select a case of low tensile strength
$S_\mathrm{max}=10^7$ erg cm$^{-3}$, which is appropriate for composite grains and
grain mantles.
We show the results in Fig.~\ref{fig:size}a. We observe that, even at $t=0.1$~Gyr, when the
grain production is dominated by stellar sources, small grains are efficiently formed by
rotational disruption.
For comparison, we also show in Fig.\ \ref{fig:size}b the evolution of grain size distribution without
rotational disruption but including shattering following HM20.
Since the efficiency of shattering depends on the dust abundance, the small-grain production
is not prominent in the early stage ($t=0.1$~Gyr). Thus, very different grain size distributions
are predicted between rotational disruption and shattering in the early phase of galaxy evolution.
At $t=0.3$ Gyr, the bump at $a\sim 0.002~\micron$ is caused by
accretion, and is seen in both panels. After $t\sim 3$ Gyr, the grain size
distributions become smooth power-law-like forms, but there are some
differences between rotational disruption and shattering.
In the case with rotational disruption, there is a strong cut-off at
$a=a_\mathrm{disr}$ (note that $a_\mathrm{disr}=0.18~\micron$ for the current case),
because the time-scale of rotational disruption is short.
At $a<a_\mathrm{disr}$, the grain size distribution is converged to a power-law-like
shape but the overall slope is different: the scenario with rotational disruption predicts
more large grains than that with shattering, since rotational disruption only occurs
at $a>a_\mathrm{disr}$. This is in contrast with shattering, in which small grains could
also be fragmented if they collide with large grains. 
As argued in HM20, the strong fragmentation cascade
by shattering together with the grain growth by coagulation leads to
an MRN-like grain size distribution with an overall slope of $p\simeq 3.5$
($n\propto a^{-p}$)\footnote{Note that the slope is $(-p+4)$ in the figure.}
\citep[see also][]{Dohnanyi:1969aa,Williams:1994aa,Tanaka:1996aa,Kobayashi:2010aa}.
However, rotational disruption cannot achieve this balance between fragmentation and coagulation
because it does not occur at small grain radii.
Less abundant small grains in rotational disruption lead to an overall slope $p$ significantly
smaller than 3.5.

\begin{figure}
\includegraphics[width=0.45\textwidth]{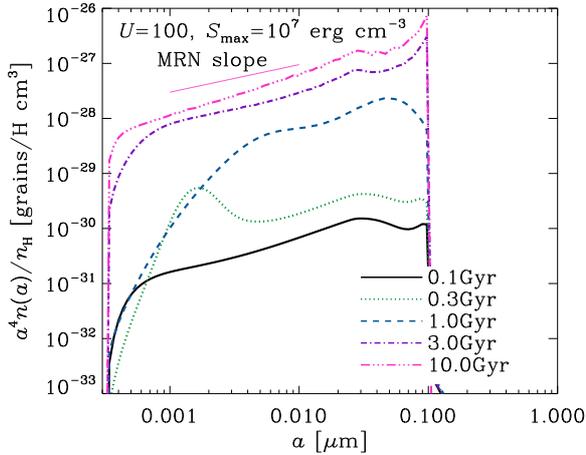}
\caption{Same as Fig.\ \ref{fig:size}a but for $U=100$.
\label{fig:size_U}}
\end{figure}

We also examine the dependence on the parameters relevant for
rotational disruption, that is, $U$ and $S_\mathrm{max}$.
First, we examine the effect of stronger radiation field by adopting
$U=100$ and $S_\mathrm{max}=10^7$ erg cm$^{-3}$. This case
is expected to predict fast disruption with a small $a_\mathrm{disr}(\simeq 0.1~\micron )$.
We observe in Fig.\ \ref{fig:size_U} [compared with Fig.\ \ref{fig:size}(a)] that,
if $U$ is larger, the grain size distribution at $a\lesssim 0.003~\micron$
is slightly more enhanced at $t\lesssim 0.3$ Gyr because of more efficient rotational disruption and
accretion (which occurs predominantly at small grain radii).
The enhancement of small grains in $U=100$ leads to
more efficient coagulation at $t=1$ Gyr
(as observed around $a\sim 0.05~\micron$).
The grain size distribution at $t\ge 3$ Gyr does not strongly
depend on $U$, except for the smaller cut-off radius
($a_\mathrm{disr}$) in the case of $U=100$.
Overall, the effects of a stronger ISRF appear in
slight enhancement of small-grain production at early ages and a cut-off at smaller
grain radii at all ages.

\begin{figure}
\includegraphics[width=0.45\textwidth]{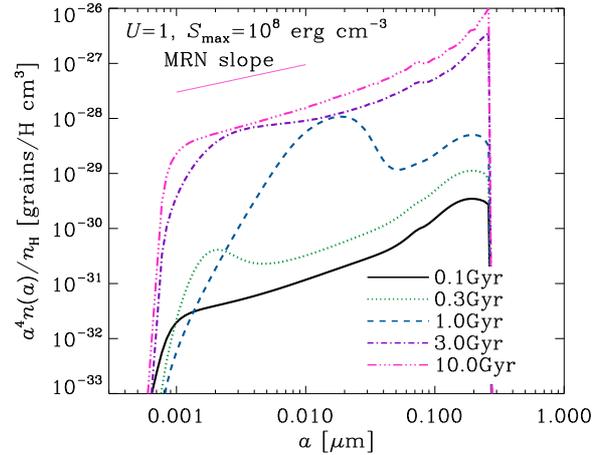}
\caption{Same as Fig.\ \ref{fig:size} but for $S_\mathrm{max}=10^8$
erg cm$^{-3}$.
\label{fig:size_Smax}}
\end{figure}

Next, we examine the dependence on $S_\mathrm{max}$. The above cases with
$S_\mathrm{max}=10^7$ erg cm$^{-3}$ represent composite grains (or grain mantles) which are
easy to disrupt. Here, we examine the cases for
$S_\mathrm{max}=10^8$ erg cm$^{-3}$ with $U=1$
($a_\mathrm{disr}=0.28~\micron$). This case is
aimed at examining relatively inefficient rotational disruption.
Note that if $S_\mathrm{max}$ is larger
than $10^8$ erg cm$^{-3}$, equation (\ref{eq:a_disr}) is not applicable, since
$a_\mathrm{disr}$ is larger than 0.28 $\micron$.  This means that radiative torques
cannot produce a grain rotation fast enough for rotational disruption.
We observe in Fig.\ \ref{fig:size_Smax} that the effect of $S_\mathrm{max}$ clearly
appears in the maximum grain radii.
The difference in $S_\mathrm{max}$ is already apparent in the grain size distributions
at $t\lesssim 0.3$ Gyr: small-grain
production is less enhanced for larger $S_\mathrm{max}$ because of less efficient
rotational disruption. At $t\sim 1$~Gyr, the bump at $a\sim 0.01$--0.03~$\micron$ is lower for
large $S_\mathrm{max}$ because less enhanced small-grain production leads to less efficient
accretion. At later stages ($t\gtrsim 3$ Gyr), the grain mass distribution ($\propto a^4 n$)
strongly peaks at $a=a_\mathrm{disr}$. Therefore, rotational disruption not only determines
the maximum grain radius but also govern the typical grain size at later ages.

\subsection{Extinction curves}\label{subsec:ext_result}

\begin{figure}
\includegraphics[width=0.48\textwidth]{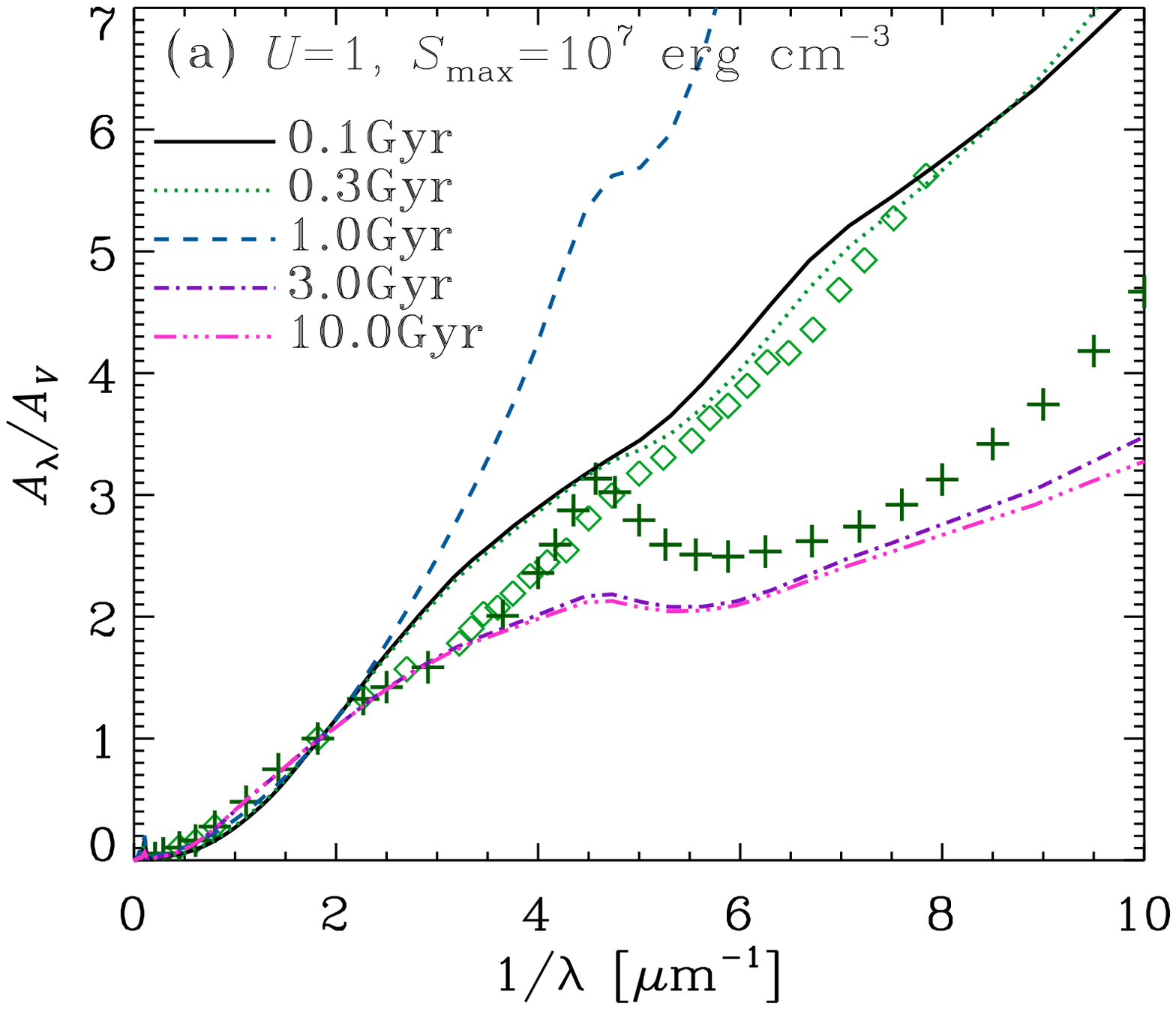}
\includegraphics[width=0.48\textwidth]{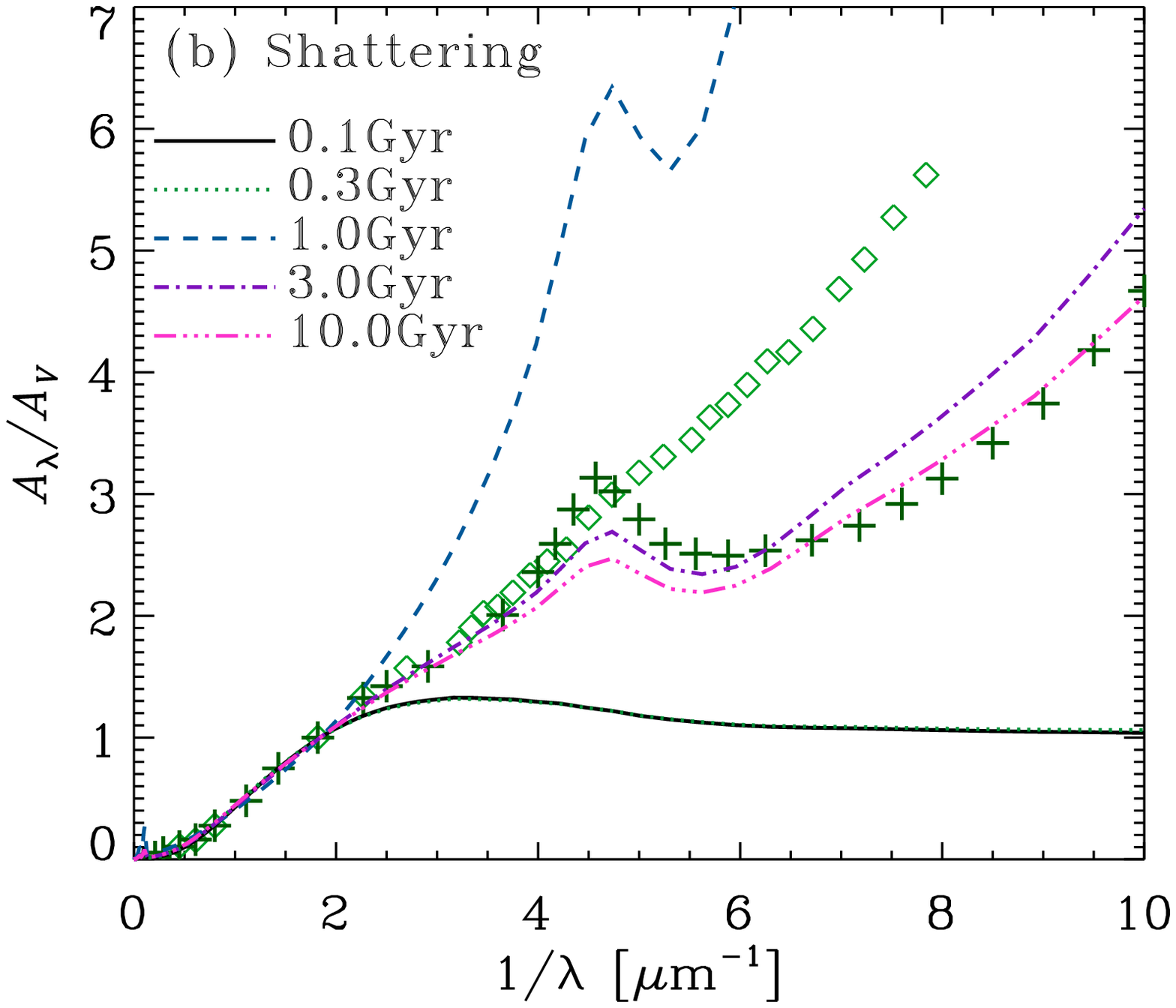}
\caption{
Extinction curves corresponding to the two cases shown
in Fig.~\ref{fig:size}. Panels (a) and (b) show the results with
rotational disruption and with shattering, respectively. The solid, dotted, dashed, dot--dashed,
and triple-dot--dashed lines show the results at
$t=0.1$, 0.3, 1, 3, and 10 Gyr, respectively.
The lines at $t=0.1$ and 0.3 Gyr are
indistinguishable in panel (b). The crosses and diamonds show the
observational data of the Milky Way and SMC extinction
curves, respectively, taken from \citet{Pei:1992aa}.
\label{fig:ext}}
\end{figure}

Based on the grain size distributions shown above, we calculate the extinction
curves by the method explained in Section \ref{subsec:ext_method}.
In Fig.~\ref{fig:ext}, we present the extinction curves corresponding to the
grain size distributions shown in Fig.\ \ref{fig:size}.
We find that the extinction curve is overall steep in the case with rotational
disruption in the earliest phases ($t\lesssim 0.3$ Gyr)
because small grains are produced by rotational disruption on a short time-scale.
Since the dust abundance is dominated by silicate, the 2175 \AA\ bump is not
prominent in the early stages. As a consequence, the extinction curve is similar
to the Small Magellanic Cloud (SMC) extinction curve with rotational disruption.
In contrast, the extinction curve is flat in the early epochs
($t\lesssim 0.3$ Gyr) with shattering,
since shattering is not efficient enough to produce small grains when the
dust abundance is small.
In both cases, the extinction curve is the steepest at $t\sim 1$ Gyr, which corresponds to the
epoch when accretion rapidly increases the abundance of small grains.
The strength of the 2175 \AA\ bump is weaker with rotational disruption than with shattering
because the small-grain abundance is lower as seen in Fig.\ \ref{fig:size}.
At later epochs ($t\geq 3$ Gyr),
the extinction curve is too flat to reproduce the Milky Way curve in the case with rotational
disruption, while it is similar to the Milky Way curve in the case with shattering.
This is because, as mentioned above, rotational disruption is less efficient at producing
very small grains than shattering.

\begin{figure}
\includegraphics[width=0.48\textwidth]{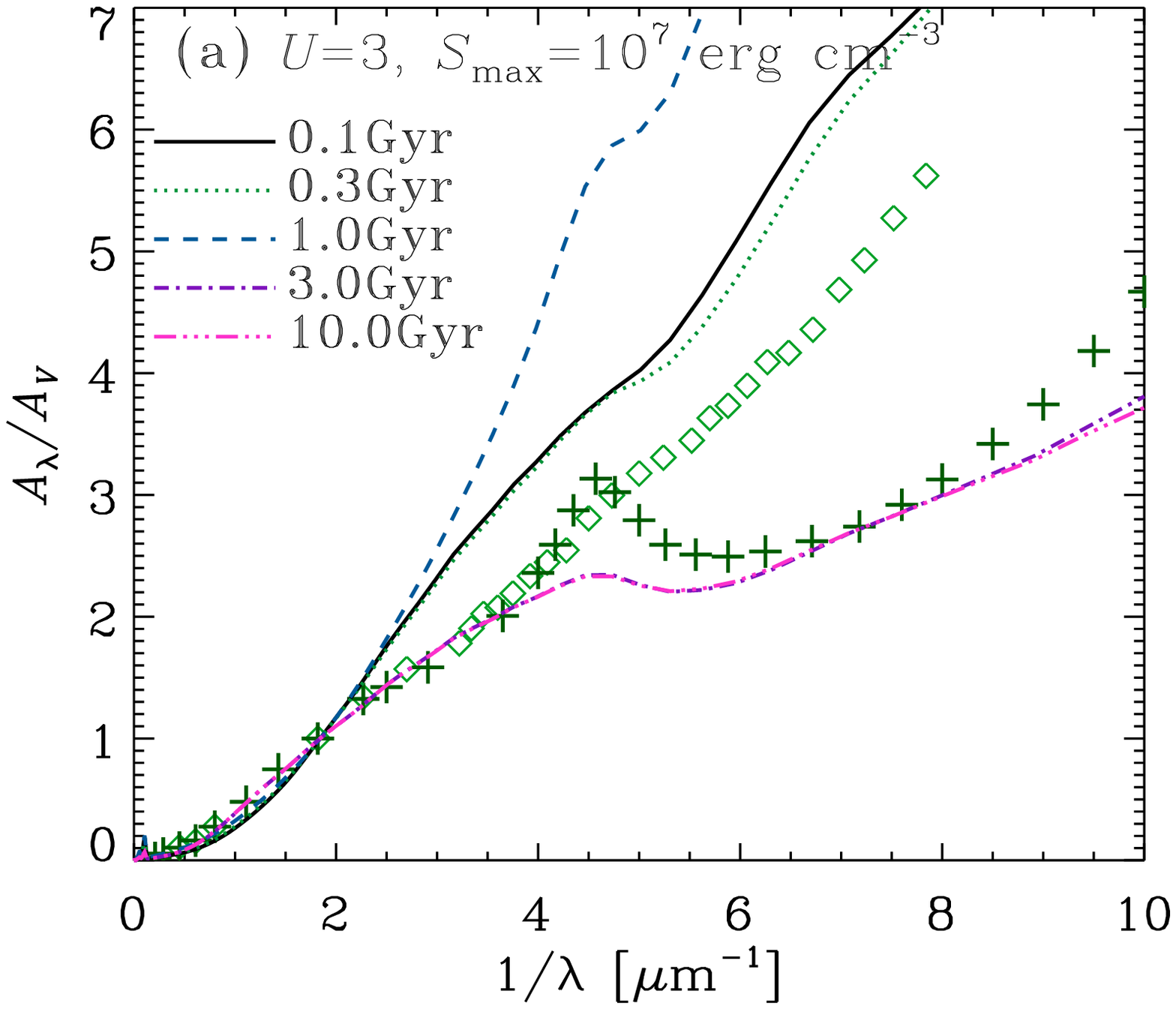}
\includegraphics[width=0.48\textwidth]{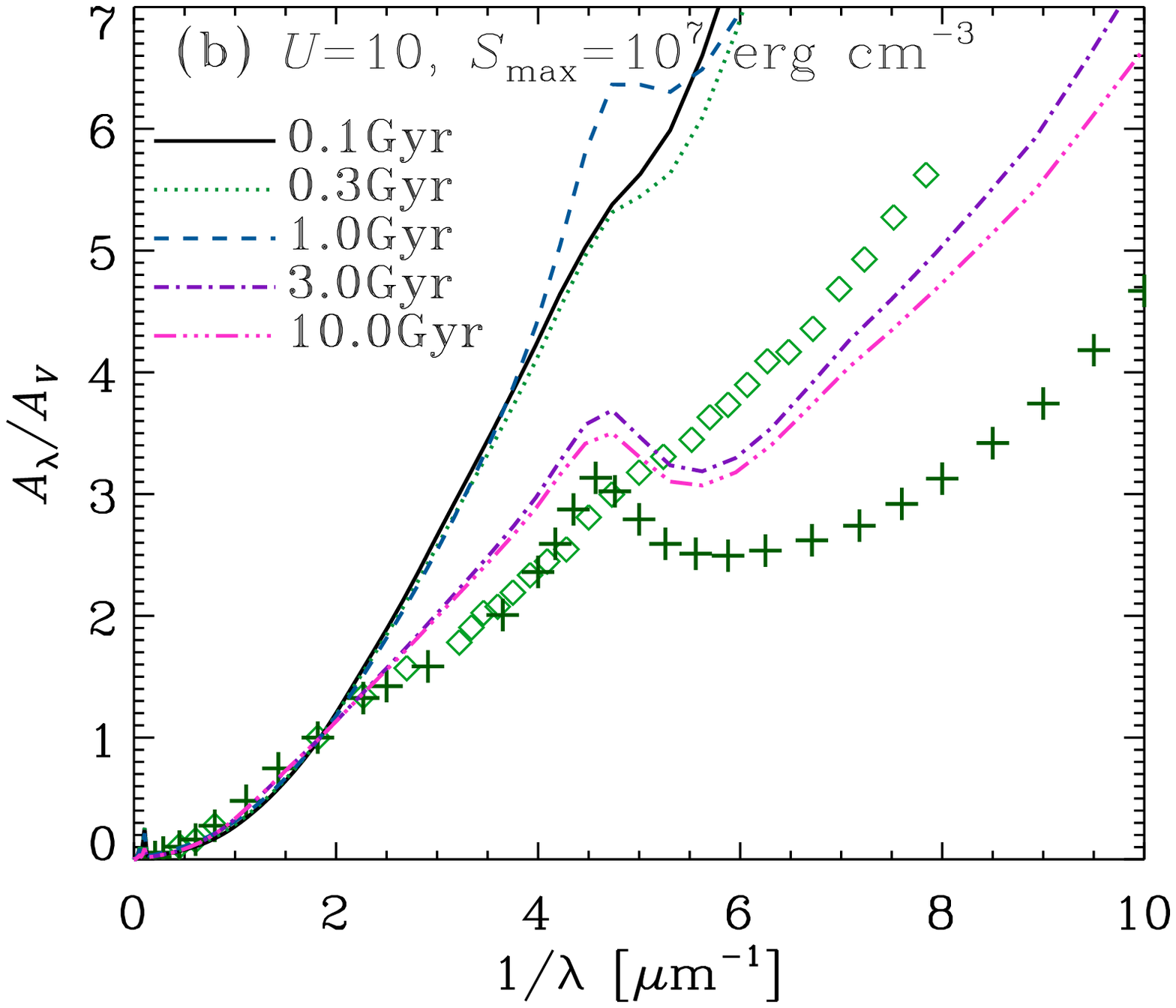}
\caption{Same as Fig.\ \ref{fig:ext}a but for higher ISRF intensities
[(a) $U=3$ and (b) $U=10$
(with $S_\mathrm{max}=10^7$ erg cm$^{-3}$)].
\label{fig:ext_param}}
\end{figure}

The effects of $U$ and $S_\mathrm{max}$ on the extinction curve are also examined.
In Fig.\ \ref{fig:ext_param}, we show the extinction curves for higher $U$
(with $S_\mathrm{max}=10^7$ erg cm$^{-3}$).
Since we found that the extinction curves are sensitive to $U$, we examine
$U=3$ and 10 here.
With stronger ISRF intensities, the grain size distributions are steeper in the early
stage of the evolution, even much steeper than the SMC extinction curve.
This is mainly because of smaller cut-off
radii for higher $U$ ($a_\mathrm{disr}=0.16$ and 0.14 $\micron$ for $U=3$ and 10, respectively).
Thus, if a galaxy has a higher ISRF intensity, we expect that the
extinction curve is steeper in the early phase of galaxy evolution.
At later epochs, the extinction curve is flattened; however, it is difficult to reproduce the
Milky Way extinction curve. For $U=3$, the carbon bump and UV slope are less prominent
than the observed Milky Way curve. On the other hand, if we increase $U$ further (as seen
in the case of $U=10$), the extinction
curve become steeper, but they do not fit the Milky Way extinction curve because of the
lack of large grains (i.e.\ $a_\mathrm{disr}$ is significantly smaller than the upper grain radius
in MRN, $0.25~\micron$). As shown by \citet{Nozawa:2013aa}, the slope and the maximum
grain radius of grain size distribution are both important to reproduce the Milky Way extinction curve.

\begin{figure}
\includegraphics[width=0.48\textwidth]{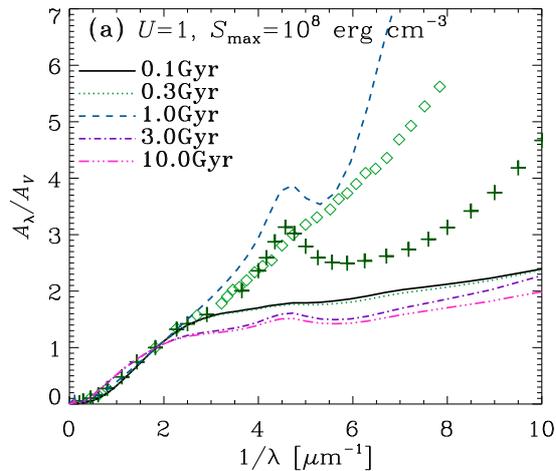}
\caption{Same as Fig.\ \ref{fig:ext}a but for a higher tensile strength
($S_\mathrm{max}=10^8$ erg cm$^{-3}$ with $U=1$).
\label{fig:ext_S}}
\end{figure}

We also show the extinction curves for higher $S_\mathrm{max}(=10^8~\mathrm{erg~cm}^{-3})$
with $U=1$ in Fig.\ \ref{fig:ext_S}
(the corresponding grain size distributions are shown in Fig.~\ref{fig:size_Smax}).
With a higher tensile strength, the extinction curves are flat in the early stage of
galaxy evolution because of larger $a_\mathrm{disr}(=0.28~\micron )$.
Note that the steep extinction curve at $t=1$~Gyr is due to accretion. Except for this short
phase of efficient accretion, the extinction curves are generally flat for
$S_\mathrm{max}\geq 10^8$ erg cm$^{-3}$. In other words, in order to produce extinction curves
as steep as the SMC curves with
rotational disruption, the tensile strength of large grains should be as small as
expected for composites and grain mantles ($S_\mathrm{max}\sim 10^7$ erg cm$^{-3}$). 

\section{Discussion}\label{sec:discussion}

\subsection{Rotational disruption compared with shattering}\label{subsec:importance}

The time-scale of rotational disruption is typically shorter than that of
chemical (dust) enrichment. Moreover,
unlike shattering, the time-scale of rotational disruption is independent of
the dust abundance. Therefore,
the effect of rotational disruption appears even in the early stage of
galaxy evolution.
If the tensile strength is as low as expected for composites and grain mantles
($S_\mathrm{max}\sim 10^7$ erg cm$^{-3}$),
the threshold grain radius for rotational disruption is less than 0.2~$\micron$ even
in the Milky Way ISRF ($U\sim 1$).
Even though the dust supplied from stellar sources is dominated by large
grains, efficient production of small grains by rotational disruption makes the grain
size distribution completely different from the original log-normal shape
(Fig.\ \ref{fig:size}).
Thus, rotational disruption provides a possible way of
steepening extinction curves in the early stage of galaxy evolution.
It is interesting to point out that, if rotational disruption is efficient,
the SMC extinction curve is reproduced (Fig.\ \ref{fig:ext}).
This is because the extinction curve becomes steep
even in the early epoch when the carbon fraction is still low (note that
graphite produces a prominent bump at 2175 \AA).

{Although the above statement for the early phase of galaxy evolution is qualitatively robust,
the extinction curves qualitatively depends on the fragment grain size distribution.
As commented in Section \ref{subsec:rot}, the assumption that the fragment grain size
distribution is similar to the case of shattering may be optimistic in the production of small grains,
although it could be justified if the grain structures are strongly inhomogeneous (with grain mantles,
grain monomers, etc.). However, if we are interested in the extinction curves down to
$\lambda\sim 0.1~\micron$, the above conclusions are not greatly affected as long as
fragments as small as $a\sim\lambda /(2\upi )\sim 0.016~\micron$ are formed.
If the minimum fragment size is larger than this, the extinction curves are not as steep as
predicted in this paper (but are still steeper than the case without rotational disruption).}

In later ($t\gtrsim 3$ Gyr) epochs, the extinction curves are rather flat
since rotational disruption only fragments grains at $a>a_\mathrm{disr}$.
Therefore, the resulting grain size distribution has a value of $p$
(characteristic slope) smaller than the MRN value (3.5).
Even if we increase $U$ to decrease $a_\mathrm{disr}$, the Milky Way extinction curve is
hard to reproduce because of $p$ significantly smaller than 3.5.

As shown above, rotational disruption affects the evolution of grain size distribution
even with $U=1$ if the tensile strength is as small as
expected for composites or grain mantles ($S_\mathrm{max}\sim 10^7$ erg cm$^{-3}$).
For example, if large grains are formed by coagulation, it is expected that the grains are
composed of many monomers, some of which may be bound loosely.
However,  if large grains originate from condensation in stellar ejecta, they likely
have a compact structure. The requirement for $S_\mathrm{max}$
is relaxed in high ISRF fields as shown later; that is, rotational disruption could
efficiently produce small grains for a wide range of tensile strengths (or grain structures).
Therefore, in the future, we need to develop a framework that treats the evolution of the
following three physical components consistently:
grain material properties (especially tensile strength), ISRF, and grain size distribution.

\subsection{Robustness of the resulting grain size distributions}\label{subsec:robustness}

In the above, we have shown that, at later ages, the dust abundance is dominated by
grains around
$a\sim a_\mathrm{disr}$ and that the overall slope $p$ is smaller than 3.5.
Here we examine if these characteristics are robust against the uncertainties
in the model. Since the overall slope is determined by the balance between
small- and large-grain productions by rotational disruption and coagulation,
respectively, we change some parameters that could largely change these processes.
We fix $U=1$ and $S_\mathrm{max}=10^7$ erg cm$^{-3}$.

\begin{figure}
\includegraphics[width=0.48\textwidth]{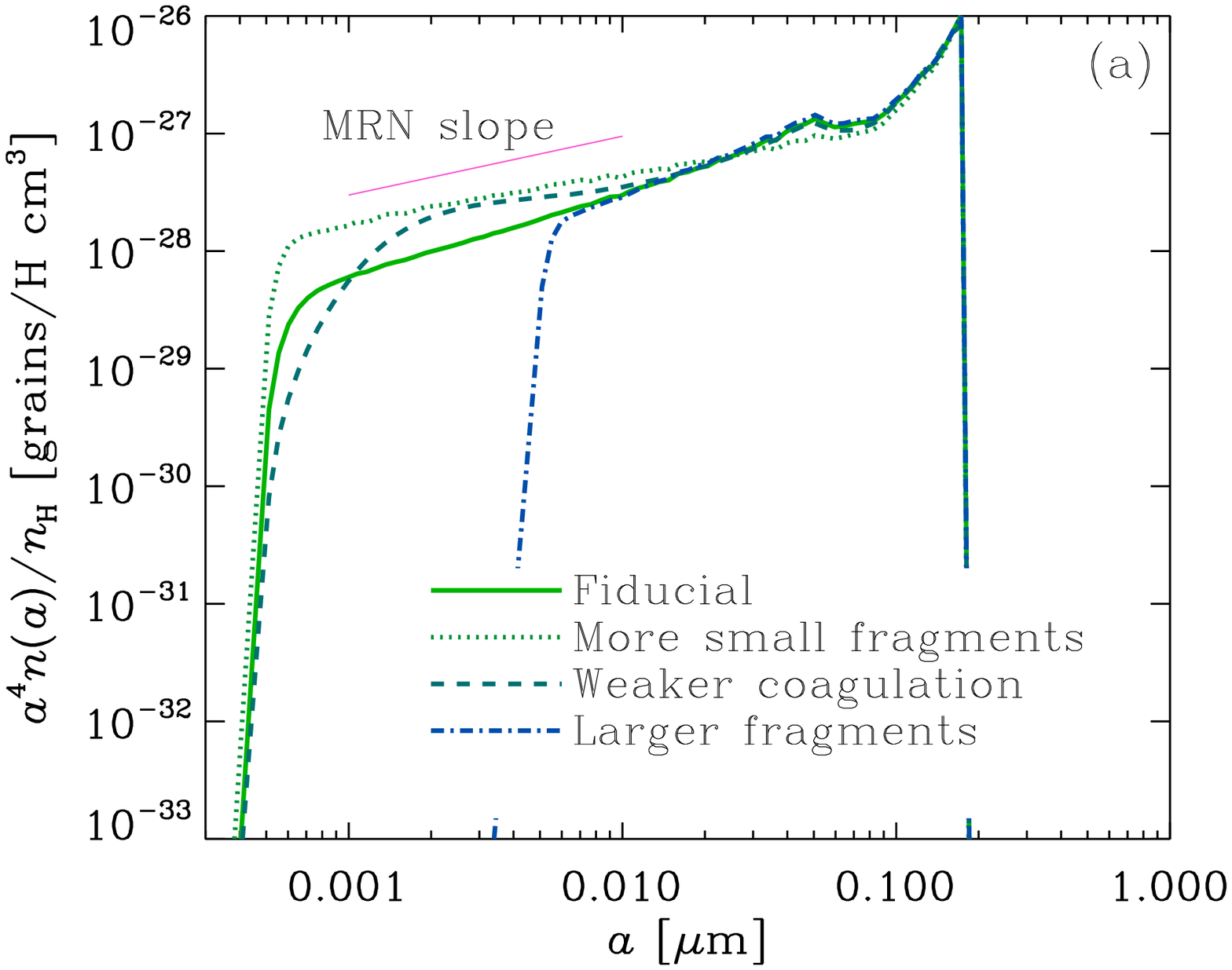}
\includegraphics[width=0.48\textwidth]{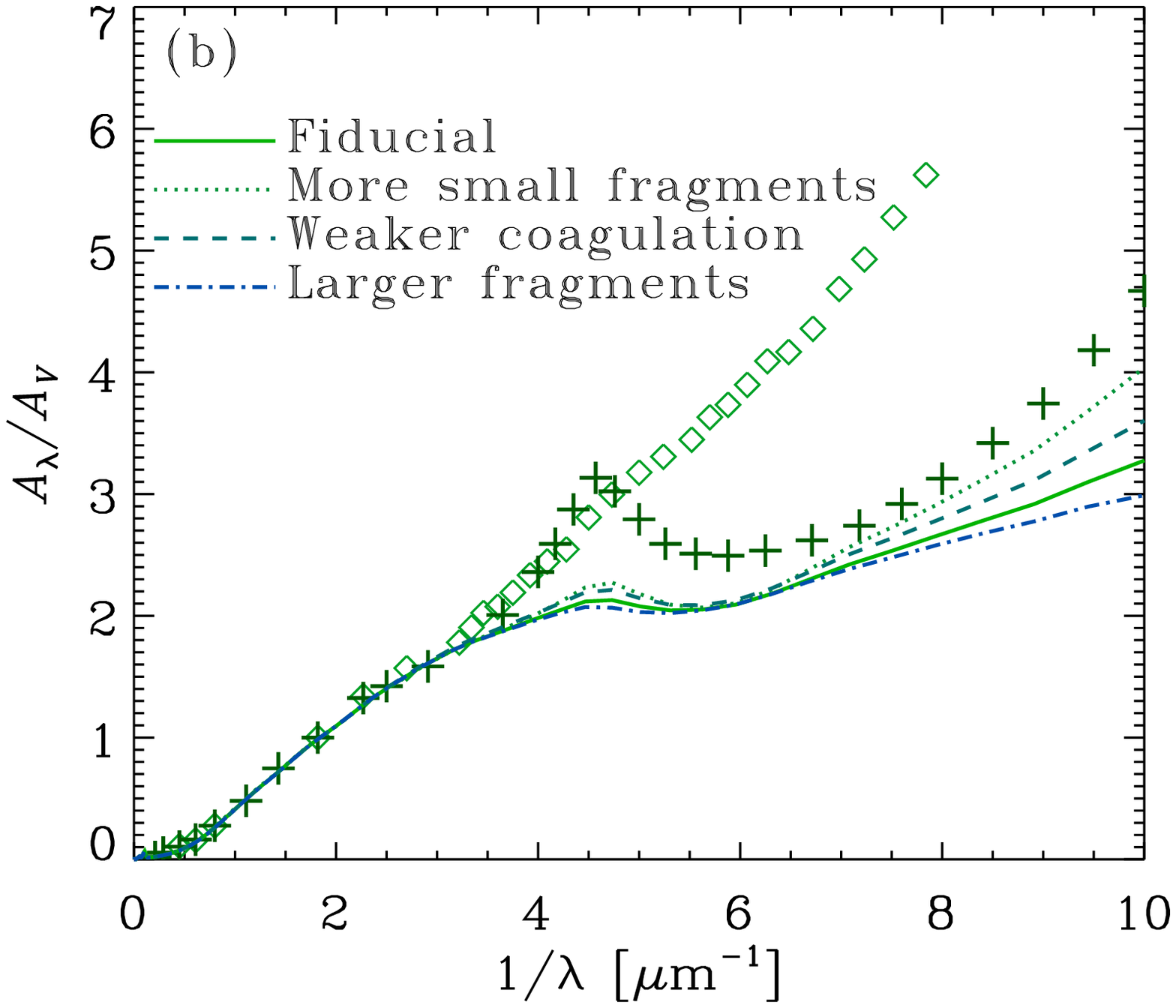}
\caption{(a) Grain size distributions at $t=10$ Gyr for different efficiencies of
small-grain production by rotational disruption and large-grain production by
coagulation. The fiducial
case (i.e.\ the same case as in Fig.\ \ref{fig:size}a), the case with $\alpha_\mathrm{f}=4$
(more small fragments in rotational disruption), the case with weaker coagulation
(with a sticking efficiency of 0.1),
{and the case with a  larger minimum fragment radius (10 times larger than the
other cases)} are shown by the solid, dotted, dashed, {and dot--dashed} lines, respectively.
The thin short solid line shows the MRN slope for reference.
(b) Extinction curves corresponding to panel (a). The line species are the same
as in panel (a), and the Milky Way (crosses) and SMC (diamonds) extinction curves are shown for
reference.
\label{fig:comp}}
\end{figure}

First, we modify rotational disruption in such a way that small-grain production is
more efficient. This is most efficiently done in our framework by changing the slope of
the fragment grain size distribution ($\alpha_\mathrm{f}$ in equation \ref{eq:frag}).
We change it to $\alpha_\mathrm{f}=4$, in which case the fragments have
an equal mass in logarithmic grain radius intervals. This value of $\alpha_\mathrm{f}$
is well out of the range suggested by \citet{Jones:1996aa} ($\alpha_\mathrm{f}=3.0$--3.4),
so that this choice of $\alpha_\mathrm{f}$ provides an extreme case.
Since the difference appears most prominently
at later ages, we only show the result at $t=10$ Gyr.
The grain size distributions and the extinction curves are compared
between the cases with $\alpha_\mathrm{f}=3.3$ (i.e.\ the above calculation)
and 4 in Fig.\ \ref{fig:comp}.

We observe in Fig.\ \ref{fig:comp}a (the solid and dotted lines)
that the grain size distributions at $a\gtrsim 0.03~\micron$ is robust
against the change of $\alpha_\mathrm{f}$.
The effect of $\alpha_\mathrm{f}$ indeed appears at small grain radii.
However, we should note that $\alpha_\mathrm{f}=4$ provides
an extreme case in terms of small-grain production as mentioned above.
{Thus, it is worth noting that the grain mass distribution has a sharp rise just below
the maximum radius ($a_\mathrm{disr}$) even in this extreme case.}

Since the accumulation of large grains could also be affected by coagulation,
we also examine the case where the coagulation efficiency is reduced by lowering
the sticking coefficient to 0.1 (we assumed unity above).
In Fig.\ \ref{fig:comp}a, we observe that the weaker coagulation indeed enhances the
grain abundance at $a\sim 0.001$--0.01~$\micron$ but that the grain
size distribution at $a\gtrsim 0.03~\micron$ is little affected by the efficiency of
coagulation. Therefore, the accumulation of grains at $a$ just below $a_\mathrm{disr}$
is robust.

{As mentioned in Section \ref{subsec:rot}, our fragment distribution function may be
optimistic in the abundance of small grains. We emphasize that even with this optimistic
assumption, the dust mass is accumulated at grain radii just below $a_\mathrm{disr}$.
However, it is worth examining a less optimistic case, where the grains are fragmented into
a much smaller number of pieces. In our model, this is easily done by making the minimum
fragment mass larger. Here we adopt $m_\mathrm{f,min}=10^{-3}m_\mathrm{f,max}$
instead of $m_\mathrm{f,min}=10^{-6}m_\mathrm{f,max}$; that is, the minimum fragment size is
10 times larger than adopted in the above calculations. The grain size distribution at $t=10$ Gyr is
shown in Fig.\ \ref{fig:comp}. We observe that the grain size distribution is simply truncated at
the smallest fragment radius ($a\sim 0.005~\micron$). Thus, we confirm that
the minimum fragment size determines the minimum grain radius in our scenario.}

We also show the extinction curves corresponding to the above grain size distributions
in Fig.\ \ref{fig:comp}b. We observe that the extinction curves are not sensitive to
$\alpha_\mathrm{f}$ or the coagulation efficiency at $\lambda\gtrsim 0.15~\micron$.
The case with
$\alpha_\mathrm{f}=4$ shows a significant steepening at far-UV because of the
excess of grains
at $a<0.01$--$0.02~\micron$ contributes to an excess of dust opacity at
$\lambda <2\upi a$.
Therefore, the extinction curve shape at $\lambda\gtrsim 0.15~\micron$ is robust
against the enhancement of fragment production and the reduction of coagulation.
{It is natural that the extinction curve is flatter for the case with the larger $m_\mathrm{f,min}$.
However, the difference is not large since the minimum grain radius (0.005 $\micron$) is still
below $\lambda /(2\pi )$. If rotational disruption only produces grains larger than 0.016 $\micron$,
the extinction curve is expected to be flattened significantly at $\lambda\sim 0.1~\micron$.}

In summary, the scenario in which small grains are predominantly
produced by rotational disruption robustly predicts that the dust abundance is dominated
by grains with $a\sim a_\mathrm{disr}$ and that the overall slope of grain size distribution
$p$ is smaller than 3.5 at later ages. These properties of grain size distribution produce
flat extinction curves; thus, it is difficult to explain the Milky Way extinction curve with
a prominent carbon bump and a steep UV slope. Probably, shattering, which is turned off
in our calculations, is still needed. However, this
does not mean that rotational disruption is unimportant. We further discuss the importance
of rotational disruption in the following subsections.

\subsection{Parameter space where rotational disruption is important}\label{subsec:param_space}

\begin{figure}
\begin{center}
\includegraphics[width=0.45\textwidth]{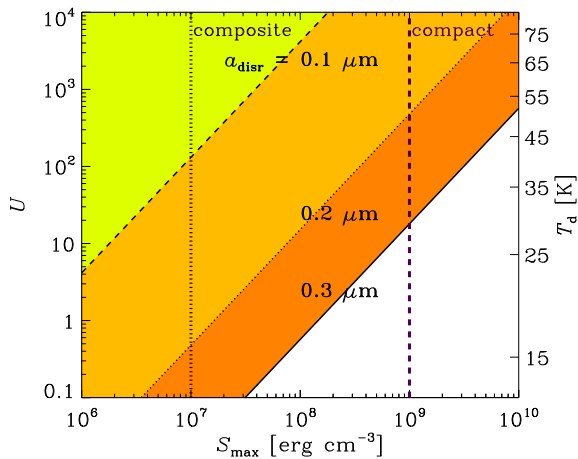}
\end{center}
\caption{Constant $a_\mathrm{disr}$ lines on the $U$--$S_\mathrm{max}$
(ISRF--tensile strength) diagram. The diagonal solid, dotted, and dashed lines show
$a_\mathrm{disr}=0.3$, 0.2, and 0.1~$\micron$, respectively.
The shaded areas present the regions where $a_\mathrm{disr}$ is smaller than
those values.
The tensile strengths discussed for composite and compact grains in the text
are shown by the vertical dotted and dashed lines, respectively. The dust temperature
corresponding to the ISRF intensity is shown on the right axis.
In the white region, rotational disruption does not occur.
\label{fig:condition}}
\end{figure}

The time-scale of rotational disruption is much shorter than that of
galaxy evolution (chemical enrichment). This means that rotational disruption
affects the grain size distribution in all the history of galaxy evolution.
The most important factor in rotational disruption is the disruption radius
$a_\mathrm{disr}$, which determines the maximum grain radius.
If we solve equation (\ref{eq:a_disr}) for $U$, we obtain
\begin{align}
U=0.48\left(\frac{a_\mathrm{disr}}{0.2~\micron}\right)^{-8.1}
\left(\frac{S_\mathrm{max}}{10^7~\mathrm{erg~cm}^{-3}}\right)^{1.5},
\end{align}
which can be regarded as an ISRF intensity that achieves a certain value of
$a_\mathrm{disr}$ under a given tensile strength.
In Fig.\ \ref{fig:condition}, we show the lines of $a_\mathrm{disr}=0.1$, 0.2, and 0.3 $\micron$
(the shaded areas show $a_\mathrm{disr}<0.1$, 0.2, and 0.3 $\micron$)
on the $U$--$S_\mathrm{max}$ plane. We also convert $U$ to the dust temperature ($T_\mathrm{d}$)
by adopting a scaling of $T_\mathrm{d}=18U^{1/6}$~K
(since the dust temperature depends on the dust composition,
we adopt the intermediate value between silicate and graphite; e.g.\ \citealt{Li:2001aa}).
As mentioned above, rotational disruption occurs only if $a_\mathrm{disr}\lesssim 0.28~\micron$
(Section \ref{subsec:rot}).
Thus, we only show $a_\mathrm{disr}$ up to 0.3 $\micron$.
This diagram gives us a clue to the significance of rotational disruption with observationally
estimated $T_\mathrm{d}$ under an assumed grain tensile strength.

In Fig.\ \ref{fig:condition}, we observe that, for a tensile strength appropriate for
composite grains ($S_\mathrm{max}\sim 10^7$ erg cm$^{-3}$), an ISRF intensity
expected for normal star-forming galaxies
(such as the Milky Way) is sufficient to destroy large ($a\gtrsim 0.2~\micron\sim$
the upper grain radius of MRN) grains.
For compact grains, which
have $S_\mathrm{max}\gtrsim 10^9$ erg cm$^{-3}$,
a strong ISRF ($U\gtrsim$ a few hundreds, or dust temperature $T_\mathrm{d}\gtrsim 50$ K)
is necessary to
disrupt grains with $a\sim 0.2~\micron$. With $U\sim 1$, rotational disruption
is not capable of disrupting compact grains. Since the maximum grain radius in the Milky Way
is $\simeq 0.25~\micron$ (MRN), rotational disruption is capable of determining the maximum
grain radius only if the grains are much softer than compact grains.

\subsection{Coexistence of rotational disruption and shattering}\label{subsec:shattering}

In this paper, in order to clarify the role of rotational disruption, we turned
off the other small-grain production mechanism -- shattering.
In fact, the shattering efficiency also depends on the tensile strength
\citep[e.g.][]{Tielens:1994aa,Jones:1996aa,Hirashita:2013ab}.
If the tensile strength is as small as expected for composite grains or grain mantles,
shattering would fragment almost all grains on the shattering time-scale
($\sim 10^8$ yr in solar-metallicity environment).
However, such a small tensile strength may be applicable to only large
grains, which are more likely to be composed of small monomers.
In this case, both shattering and rotational disruption fragment
large grains, determining the largest grain radius.
Although rotational disruption only acts on large grains ($a>a_\mathrm{disr}$),
shattering still disrupts small grains even if they are compact.
Thus, both rotational disruption and shattering affect large grains and
contribute to determining the maximum size of grains, while shattering
predominantly modifies the grain size distribution at $a<a_\mathrm{disr}$.

The largest difference between rotational disruption and shattering is as follows:
as mentioned above,
the time-scale of the former process does not depend on the dust abundance
while that of the latter does. This means that rotational disruption could
potentially be important even in the early phase of galaxy evolution.
Indeed, it has been suggested (and it is also assumed in this paper)
that dust grains formed by stellar sources
is biased to large ($a\gtrsim 0.1~\micron$) sizes. 
In the earliest phase of galaxy evolution, when dust is predominantly
supplied by stellar sources, rotational disruption could be the only mechanism
of producing small grains because shattering is inefficient in the condition of
low dust abundance. We should keep in mind that $a_\mathrm{disr}$
strongly depends on the tensile strength; therefore, it is desirable to clarify the
material properties of dust grains produced by stellar sources.

Although our knowledge on the grain-radius-dependent tensile strength
is limited, it is possible to perform the following `experiment'
to illustrate a case where rotational disruption and
shattering coexist (or `collaborate').
We simply calculate shattering based on the tensile strength appropriate for
compact grains (here we assume graphite properties as adopted in
Fig.\ \ref{fig:size}b with a tensile strength of $4\times 10^{10}$ erg cm$^{-3}$)
while we assume composite grains for rotational disruption
($S_\mathrm{max}=10^7$ erg cm$^{-3}$).
This assumption could be justified since rotational disruption only affects
large grains (which could be loosely bound as mentioned above) on a shorter
time-scale than shattering. On the other hand, small grains are more likely to be
compact so that shattering predominantly acts on compact grains.

\begin{figure}
\begin{center}
\includegraphics[width=0.48\textwidth]{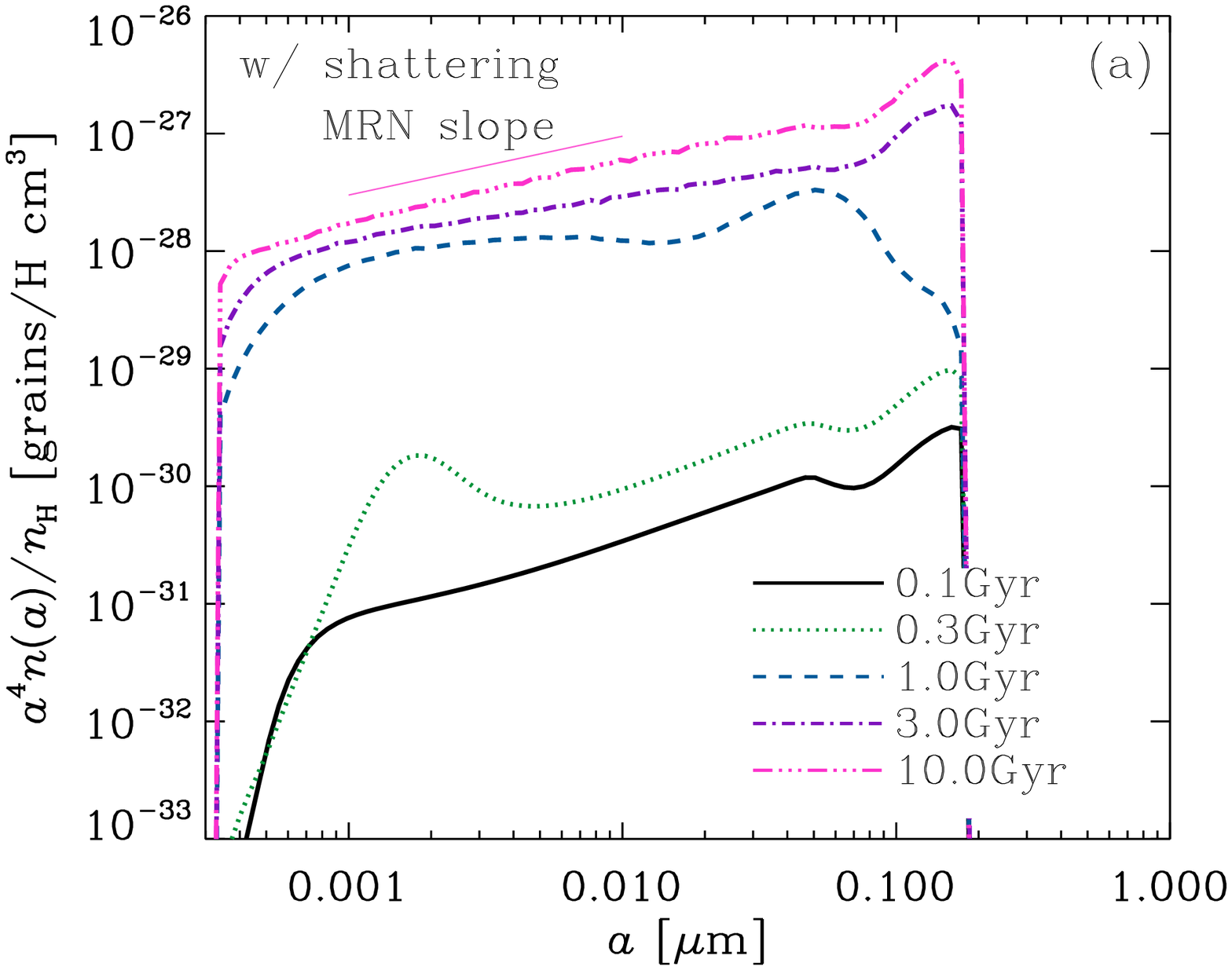}
\includegraphics[width=0.48\textwidth]{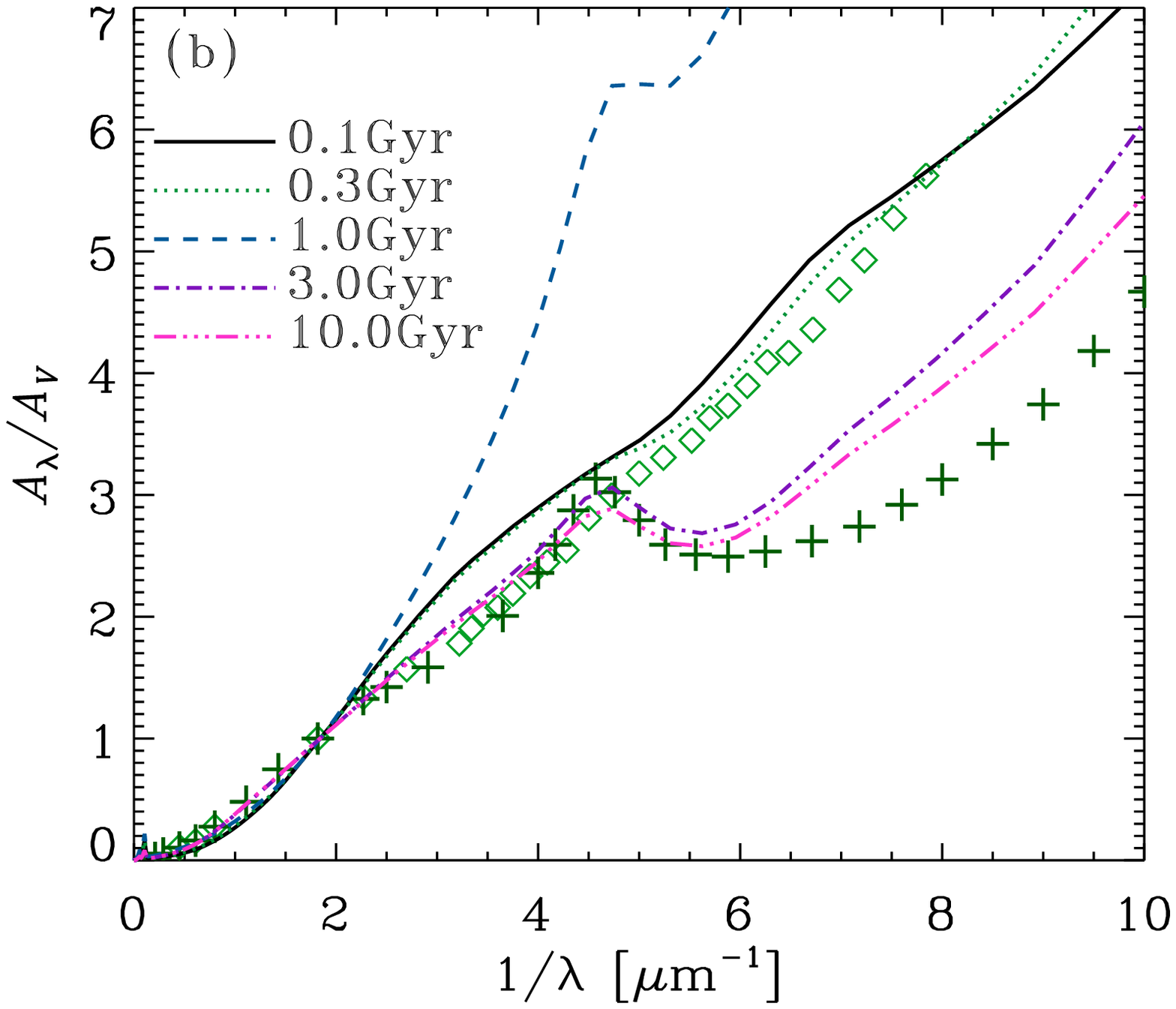}
\end{center}
\caption{(a) Evolution of grain size distribution in the condition
where rotational disruption and shattering coexist (see the text for the setting).
The solid, dotted, dashed, dot--dashed, and triple-dot--dashed lines
show the grain size distributions at $t=0.1$, 0.3, 1, 3, and 10 Gyr,
respectively. The thin dotted line shows the MRN slope.
(b) Evolution of extinction curve. The lines correspond to the same ages
as in panel (a).
The Milky Way (crosses) and SMC (diamonds) extinction curves are shown for
reference.
\label{fig:rot_shat}}
\end{figure}

In Fig.\ \ref{fig:rot_shat}(a), we show the evolution of grain size distribution
in the above setting of rotational disruption and shattering.
We observe that rotational disruption clearly determines the sharp
upper cut-off in the grain radius.
The grain size distributions at young ages ($t\lesssim 0.3$ Gyr) is similar
to the ones in Fig.\ \ref{fig:size}(a), which means that shattering is
inefficient in such young ages (because of low dust abundance).
Therefore, we confirm that rotational disruption is a dominant small-grain-production
mechanism over shattering in the early phase of galaxy evolution.
Compared again with Fig.~\ref{fig:size}(a),
at $t\gtrsim 1$ Gyr, more small grains are produced owing to shattering.
In particular, the grain size distribution eventually shows a slope consistent with
MRN ($p=3.5$). Thus, the evolution of grain size distribution with shattering
produces an MRN-like grain size distribution.
In summary, if rotational disruption and shattering coexist, the upper limit of the
grain radius is determined by rotational disruption throughout all ages, while
the slope of grain size distribution converges to the MRN value at later ages
if shattering is present.

We show the evolution of extinction curve in Fig.\ \ref{fig:rot_shat}, which is to be
compared with Fig.\ \ref{fig:ext}(a). As mentioned above, shattering is not important
in the early phase of galaxy evolution; thus, the extinction curves at $t\leq 0.3$ Gyr
is not modified by shattering. At later ages, small grains are efficiently produced by
shattering and contribute to the steepening of extinction curve.
Shattering is thus confirmed to be important
in producing extinction curves as steep as the Milky Way curve.
In summary, rotational disruption can predominantly shape the extinction curve in the
early phase of galaxy evolution, while shattering could be necessary to explain
an extinction curve as steep as the Milky Way curve at later ages.

\subsection{Implication for starburst galaxies}\label{subsec:starburst}

In the above, we have fixed the star formation time-scale $\tau_\mathrm{SF}$ to
5 Gyr.
On the other hand, the typical star formation time-scale of starburst galaxies
is a few ${}\times 10^8$ yr \citep[e.g.][]{Larson:1978aa}.
Starburst galaxies are interesting for rotational disruption because of their high ISRF
environments. Following our
previous paper, HM20, we adopt $\tau_\mathrm{SF}=0.5$ Gyr and
$\eta_\mathrm{dense}=0.9$ to describe short star-formation time-scales and
predominantly dense environments in starburst galaxies.
This is referred to as the starburst model.
We examine $U=1000$ (corresponding to $T_\mathrm{d}\sim 60$ K) for
an extreme ISRF environment actually observed in starburst galaxies
\citep[e.g.][]{Zavala:2018aa,Lim:2020aa}.
It is also interesting to examine a possibility of
destroying compact grains, so that we examine $S_\mathrm{max}=10^9$ erg cm$^{-3}$
as well as $10^7$ erg cm$^{-3}$.

\begin{figure}
\begin{center}
\includegraphics[width=0.45\textwidth]{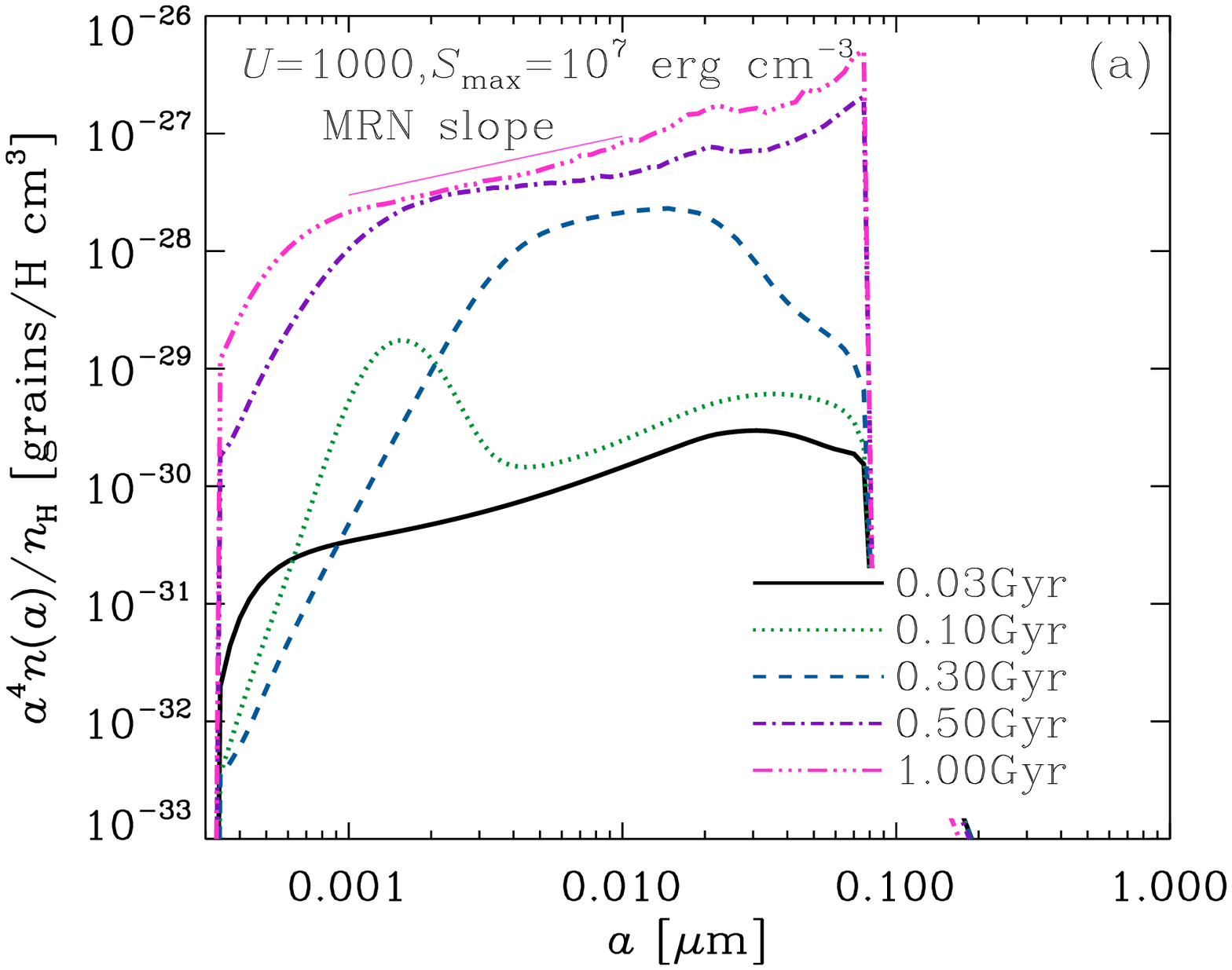}
\includegraphics[width=0.45\textwidth]{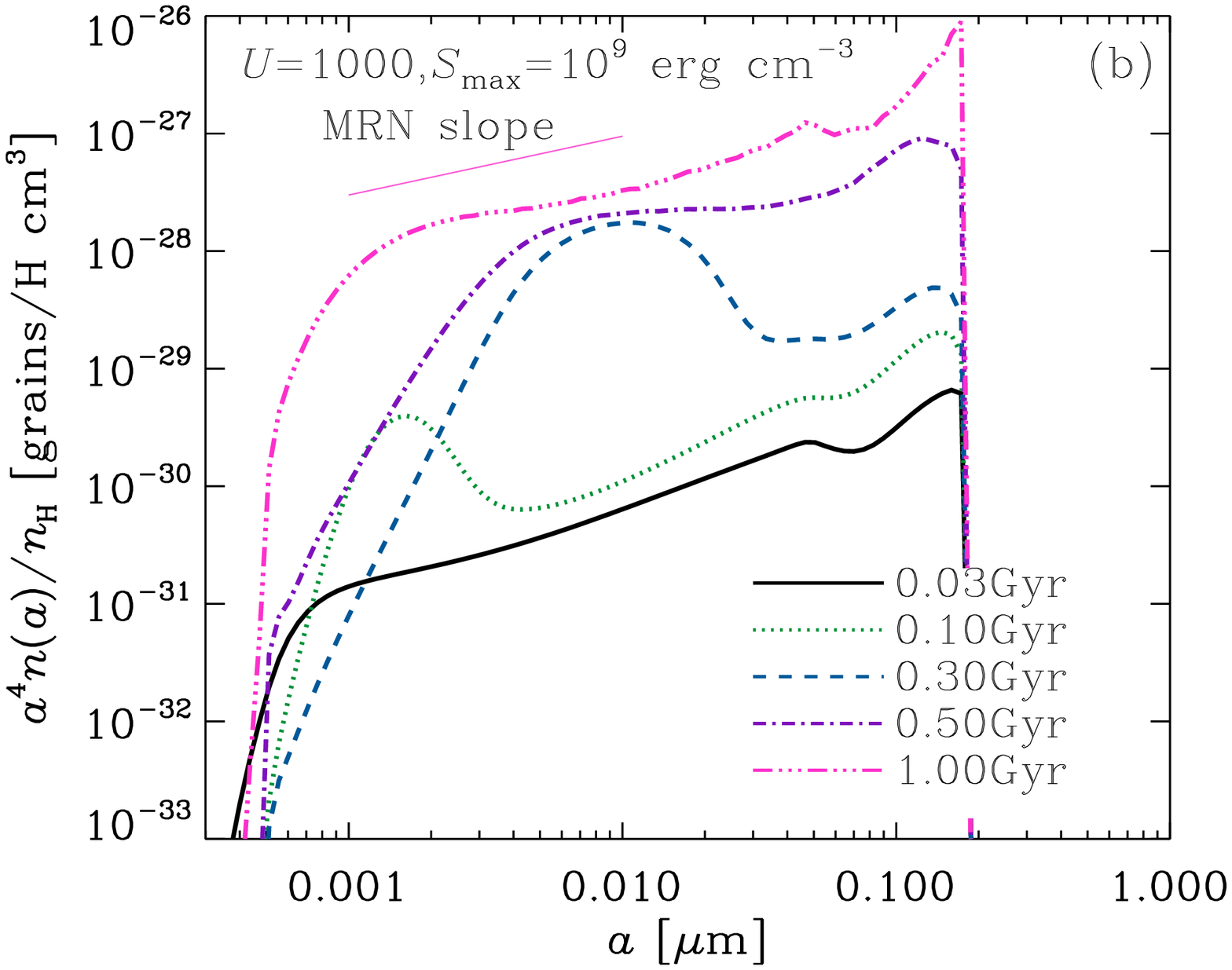}
\end{center}
\caption{Evolution of grain size distribution for the starburst model
(see the text). We adopt $S_\mathrm{max}=10^7$ and $10^9$ erg cm$^{-3}$
in panels (a) and (b), respectively, and fix $U=1000$.
The solid, dotted, dashed, dot--dashed, and triple-dot--dashed lines
show the grain size distributions at $t=0.03$, 0.1, 0.3, 0.5, and 1 Gyr,
respectively. The thin dotted line shows the MRN slope.
\label{fig:size_sb}}
\end{figure}

In Fig.\ \ref{fig:size_sb}, we show the resulting grain size
distributions. Since the evolutionary time-scale is short,
we show $t=0.03$, 0.1, 0.3, 0.5, and 1 Gyr. 
We observe that, except for the difference in the time-scale, the
evolutionary behaviour of the grain size distribution is similar to that
already investigated in Fig.\ \ref{fig:size_U}.
Small grains are abundant already at $t<0.1$ Gyr because
rotational disruption supplies small grains.
Comparing the two panels in Fig.\ \ref{fig:size_sb},
the most prominent difference appears at the cut-off radius ($a_\mathrm{disr}$).
However, the grain size distribution at $a\ll a_\mathrm{disr}$ is little affected by
the difference in $S_\mathrm{max}$.
With the high value of $U=1000$,
even in the case of $S_\mathrm{max}=10^9$ erg cm$^{-3}$,
$a_\mathrm{disr}$ becomes smaller than 0.2 $\micron$.

\begin{figure}
\begin{center}
\includegraphics[width=0.45\textwidth]{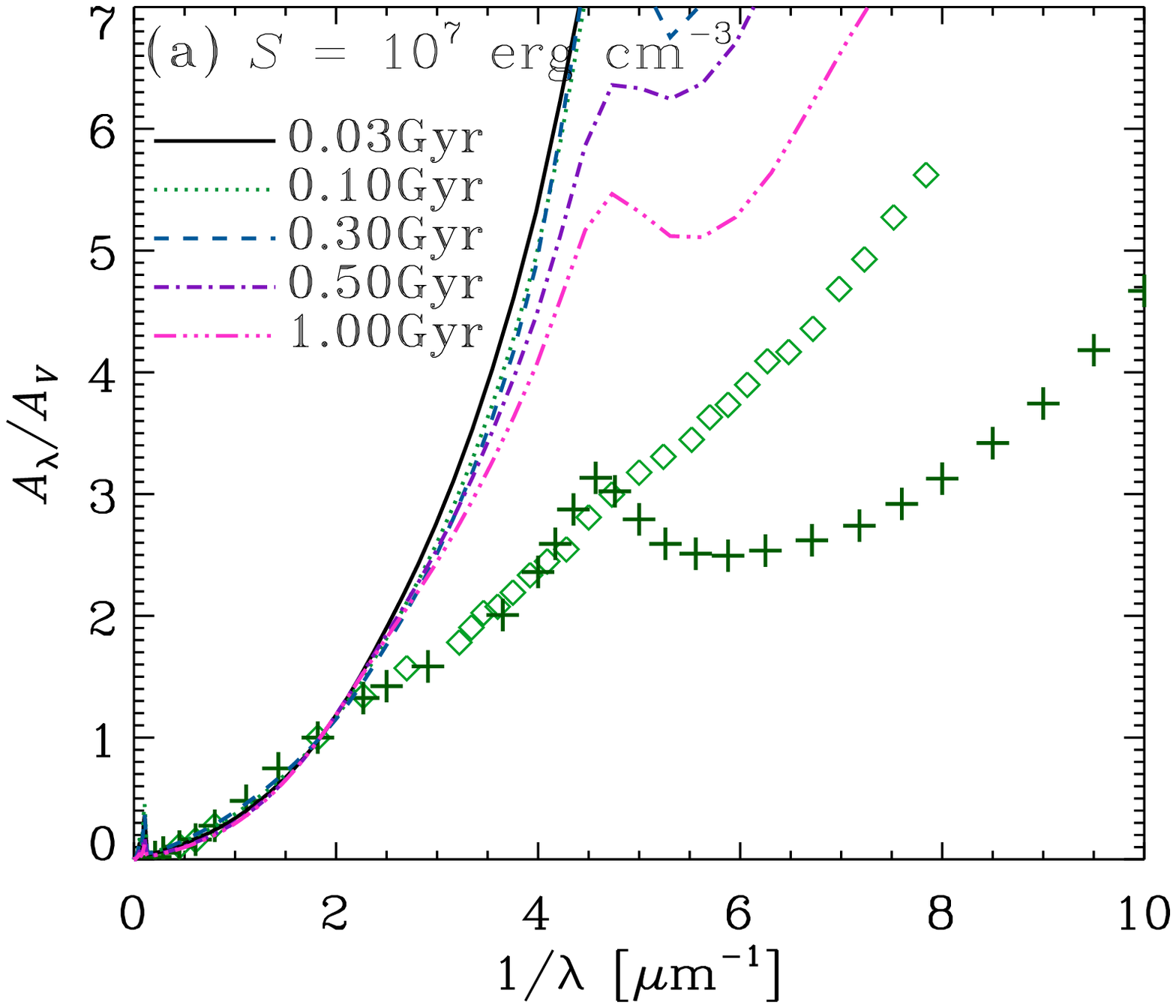}
\includegraphics[width=0.45\textwidth]{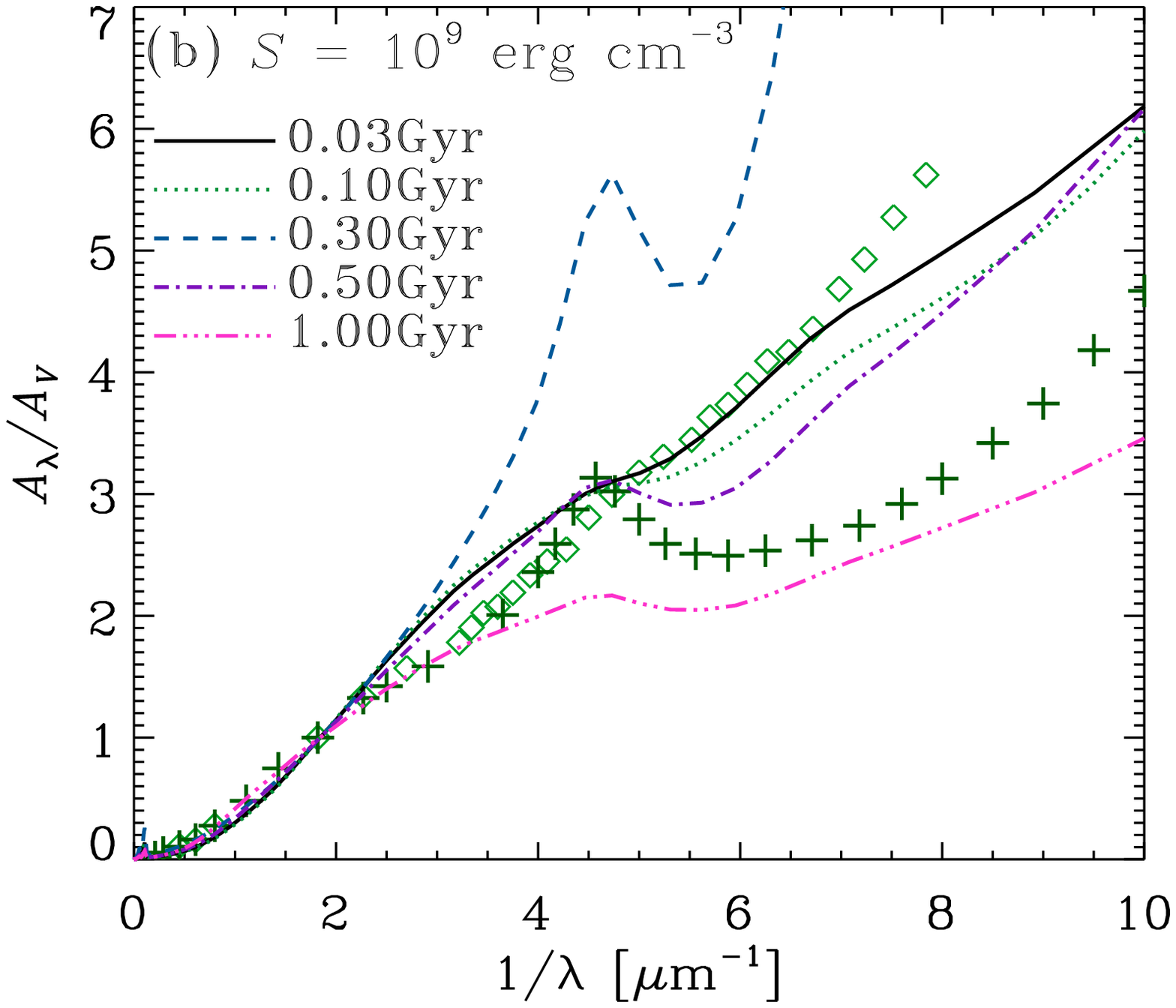}
\end{center}
\caption{Evolution of extinction curve for the starburst model (see the text).
Panels (a) and (b) present the results for the grain size distributions in the
corresponding panels in Fig.\ \ref{fig:size_sb}.
The line species for the ages are the same as in Fig.\ \ref{fig:size_sb}
(also shown in each panel).
\label{fig:ext_sb}}
\end{figure}

In Fig.\ \ref{fig:ext_sb}, we show the extinction curves corresponding to the
above grain size distributions for the starburst models.
We observe that the extinction curves stay steep for $S_\mathrm{max}=10^7$ erg cm$^{-3}$
because of the small maximum grain radius ($a_\mathrm{disr}$).
Even for the harder-grain case of $S_\mathrm{max}=10^9$ erg cm$^{-3}$,
a significant steepening of extinction curve is seen even at $t<0.1$ Gyr because of the
efficient small-grain production by rotational disruption.
It is interesting to point out that the extinction curves are similar to the SMC curve at
young ages. The extinction curves become flatter at later stages because of coagulation.
Therefore, in starburst galaxies, the small-grain production by rotational disruption could put a significant
imprint on the extinction curves especially at young ages.

For some starburst galaxies, attenuation curves, which include all the radiation transfer
effects, are obtained instead of extinction curves \citep[e.g.][]{Calzetti:2001aa}.
The effects of dust distribution
geometry and of stellar-age-dependent extinction make the attenuation curve
significantly different from the original extinction curve
\citep[e.g.][]{Witt:2000aa,Inoue:2005aa,Seon:2016aa,Narayanan:2018aa}.
Therefore, the fact that the extinction curves derived in this paper
are different from the so-called Calzetti attenuation curve is not a contradiction.
We leave radiation transfer calculations of attenuation curves
and comparison with observations for future work.

\subsection{Implication for high-redshift galaxies}
\label{subsec:highz}

The evolution of dust temperature along the redshift is still being debated.
While there have been some observational indications that the dust temperature
tends to be higher in higher redshift galaxies
\citep{Symeonidis:2013aa,Bethermin:2015aa,Schreiber:2018aa,Zavala:2018aa},
the trend could be driven by an observational bias \citep{Lim:2020aa}.
At $z\gtrsim 5$, Lyman break galaxies (LBGs) and
Lyman $\alpha$ emitters have
dust temperatures typically higher than $\sim 35$ K,
and some could have dust temperatures as high as $\gtrsim 70$ K
\citep[e.g.][]{Hirashita:2017ab,Bakx:2020aa}.
\citet{Ferrara:2017aa} also theoretically suggested that the dust temperature in the diffuse
ISM of high-redshift LBGs could be as high as 35--60 K.
These dust temperatures correspond to $U\sim$ a few tens to
a few thousands.
There are also some extreme populations of galaxies whose dust temperatures
even reach $\sim 90$ K \citep{Toba:2020aa}.
This indicates that rotational disruption could have a significant
imprint on the grain size distributions and the extinction curves in high-redshift
galaxies.

\citet{Liu:2019aa} suggested that the extinction curves in high-redshift galaxies give
us a clue to the dominant dust sources in the early Universe. Their dust evolution model
did not include rotational disruption, so it predicted that the grains are biased
to large sizes if the stellar dust production is dominant.
However, if rotational disruption is as efficient as predicted by \citet{Hoang:2019aa} and this paper,
small grains are abundant even if the dust production is dominated by stellar sources.
Therefore, the steepness of extinction curve does not solely reflect the major dust sources.
To isolate the effect of rotational disruption, it would be necessary to examine if
the steepness of extinction curve has a correlation with the dust temperature (or ISRF
intensity).

\section{Conclusion}\label{sec:conclusion}

We formulate and calculate the evolution of grain size distribution in a galaxy by
newly considering the effect of rotational disruption. We basically use the model developed
in our previous papers but include rotational disruption instead of shattering for the major
small-grain production mechanism. We evaluate the time-scale and grain radius threshold (above which
grains are disrupted) by considering the dependence on
the tensile strength of the grains and the ambient ISRF intensity
\citep{Hoang:2019aa}.

We find that rotational disruption has a large influence on the evolution of grain size distribution
in the following two aspects if the large grains are have a tensile strength
$S_\mathrm{max}\sim 10^7$ erg cm$^{-3}$ as expected for composite grains and grain
mantles. First, because of the short time-scale of rotational disruption,
the small-grain production occurs even in the early stage of galaxy evolution, when the grain
production is dominated by stellar sources. Therefore, even though stars produce large grains,
rotational disruption enhances
the abundance of small grains. {As a consequence, the extinction curves could be as steep as}
the SMC extinction curve {even in the early phase of galaxy evolution}.
Secondly, the upper bound of the grain radius is determined by the
threshold of rotational disruption ($a_\mathrm{disr}$). The steepness of extinction curve is basically
regulated by $a_\mathrm{disr}$ at later epochs when interstellar processing dominates the grain
size distribution. For compact grains with $S_\mathrm{max}\gtrsim 10^9$ erg cm$^{-3}$,
the grain size evolution is significantly affected by rotational disruption if the radiation field is as strong
as expected for starburst galaxies ($U\gtrsim$ a few hundreds or $T_\mathrm{d}\gtrsim 50$ K).

Rotational disruption differs from shattering in that it only occurs at the largest grain radii
($a>a_\mathrm{disr}$).
Therefore, we observe a concentration of grains around $a\sim a_\mathrm{disr}$,
and an overall slope shallower than the MRN value ($p<3.5$).
This produces too flat an extinction curve at later stages to reproduce the Milky Way extinction
curve unless $a_\mathrm{disr}\lesssim 0.1~\micron$
as expected in starburst galaxies. Rotational disruption tends to predict steep extinction curves for
starburst galaxies.

If the ISRF is high in high-redshift galaxies as indicated by some observations,
rotational disruption could play an important role in determining the grain size distributions in
the early Universe. 
Thus, if the steepening of extinction curves in high-redshift galaxies is correlated with
the dust temperature (or the ISRF intensity), we could argue that
rotational disruption is really acting efficiently in the early stage of galaxy evolution.

\section*{Acknowledgements}
 
{We are grateful to the referee, K. Silsbee, for useful comments.}
HH thanks the Ministry of Science and Technology for support through grants
MOST 107-2923-M-001-003-MY3 and MOST 108-2112-M-001-007-MY3
(RFBR 18-52-52006).
TH acknowledges the support by the National Research Foundation of Korea (NRF) grants funded by the Korea government (MSIT) through the Basic Science Research Program (2017R1D1A1B03035359) and Mid-career Research Program (2019R1A2C1087045).



\bibliographystyle{mnras}
\bibliography{/Users/hirashita/bibdata/hirashita}

\begin{thebibliography}{}
\makeatletter
\relax
\def\mn@urlcharsother{\let\do\@makeother \do\$\do\&\do\#\do\^\do\_\do\%\do\~}
\def\mn@doi{\begingroup\mn@urlcharsother \@ifnextchar [ {\mn@doi@}
  {\mn@doi@[]}}
\def\mn@doi@[#1]#2{\def\@tempa{#1}\ifx\@tempa\@empty \href
  {http://dx.doi.org/#2} {doi:#2}\else \href {http://dx.doi.org/#2} {#1}\fi
  \endgroup}
\def\mn@eprint#1#2{\mn@eprint@#1:#2::\@nil}
\def\mn@eprint@arXiv#1{\href {http://arxiv.org/abs/#1} {{\tt arXiv:#1}}}
\def\mn@eprint@dblp#1{\href {http://dblp.uni-trier.de/rec/bibtex/#1.xml}
  {dblp:#1}}
\def\mn@eprint@#1:#2:#3:#4\@nil{\def\@tempa {#1}\def\@tempb {#2}\def\@tempc
  {#3}\ifx \@tempc \@empty \let \@tempc \@tempb \let \@tempb \@tempa \fi \ifx
  \@tempb \@empty \def\@tempb {arXiv}\fi \@ifundefined
  {mn@eprint@\@tempb}{\@tempb:\@tempc}{\expandafter \expandafter \csname
  mn@eprint@\@tempb\endcsname \expandafter{\@tempc}}}

\bibitem[\protect\citeauthoryear{{Aoyama}, {Hirashita}  \& {Nagamine}}{{Aoyama}
  et~al.}{2020}]{Aoyama:2020aa}
{Aoyama} S.,  {Hirashita} H.,   {Nagamine} K.,  2020, \mnras, \href
  {https://ui.adsabs.harvard.edu/abs/2019arXiv190601917A} {491, 3844}

\bibitem[\protect\citeauthoryear{{Asano}, {Takeuchi}, {Hirashita}  \&
  {Nozawa}}{{Asano} et~al.}{2013}]{Asano:2013aa}
{Asano} R.~S.,  {Takeuchi} T.~T.,  {Hirashita} H.,   {Nozawa} T.,  2013,
  \mn@doi [\mnras] {10.1093/mnras/stt506}, \href
  {http://adsabs.harvard.edu/abs/2013MNRAS.432..637A} {432, 637}

\bibitem[\protect\citeauthoryear{{Bakx} et~al.,}{{Bakx}
  et~al.}{2020}]{Bakx:2020aa}
{Bakx} T. J.~L.~C.,  et~al., 2020, arXiv e-prints, \href
  {https://ui.adsabs.harvard.edu/abs/2020arXiv200102812B} {p. arXiv:2001.02812}

\bibitem[\protect\citeauthoryear{{B{\'e}thermin} et~al.,}{{B{\'e}thermin}
  et~al.}{2015}]{Bethermin:2015aa}
{B{\'e}thermin} M.,  et~al., 2015, \mn@doi [\aap]
  {10.1051/0004-6361/201425031}, \href
  {https://ui.adsabs.harvard.edu/abs/2015A&A...573A.113B} {573, A113}

\bibitem[\protect\citeauthoryear{{Bohren} \& {Huffman}}{{Bohren} \&
  {Huffman}}{1983}]{Bohren:1983aa}
{Bohren} C.~F.,  {Huffman} D.~R.,  1983, Absorption and Scattering of Light by
  Small Particles.
Wiley

\bibitem[\protect\citeauthoryear{{Burgarella} et~al.,}{{Burgarella}
  et~al.}{2013}]{Burgarella:2013aa}
{Burgarella} D.,  et~al., 2013, \mn@doi [\aap] {10.1051/0004-6361/201321651},
  \href {http://adsabs.harvard.edu/abs/2013A%26A...554A..70B} {554, A70}

\bibitem[\protect\citeauthoryear{{Calzetti}}{{Calzetti}}{2001}]{Calzetti:2001aa}
{Calzetti} D.,  2001, \mn@doi [\pasp] {10.1086/324269}, \href
  {http://adsabs.harvard.edu/abs/2001PASP..113.1449C} {113, 1449}

\bibitem[\protect\citeauthoryear{{Cazaux} \& {Tielens}}{{Cazaux} \&
  {Tielens}}{2004}]{Cazaux:2004aa}
{Cazaux} S.,  {Tielens} A.~G.~G.~M.,  2004, \mn@doi [\apj] {10.1086/381775},
  \href {http://adsabs.harvard.edu/abs/2004ApJ...604..222C} {604, 222}

\bibitem[\protect\citeauthoryear{{Chabrier}}{{Chabrier}}{2003}]{Chabrier:2003aa}
{Chabrier} G.,  2003, \mn@doi [\pasp] {10.1086/376392}, \href
  {http://adsabs.harvard.edu/abs/2003PASP..115..763C} {115, 763}

\bibitem[\protect\citeauthoryear{{Chen}, {Hirashita}, {Hou}, {Aoyama},
  {Shimizu}  \& {Nagamine}}{{Chen} et~al.}{2018}]{Chen:2018aa}
{Chen} L.-H.,  {Hirashita} H.,  {Hou} K.-C.,  {Aoyama} S.,  {Shimizu} I.,
  {Nagamine} K.,  2018, \mn@doi [\mnras] {10.1093/mnras/stx2863}, \href
  {http://adsabs.harvard.edu/abs/2018MNRAS.474.1545C} {474, 1545}

\bibitem[\protect\citeauthoryear{{Dohnanyi}}{{Dohnanyi}}{1969}]{Dohnanyi:1969aa}
{Dohnanyi} J.~S.,  1969, \mn@doi [\jgr] {10.1029/JB074i010p02531}, \href
  {https://ui.adsabs.harvard.edu/abs/1969JGR....74.2531D} {74, 2531}

\bibitem[\protect\citeauthoryear{{Draine} \& {Weingartner}}{{Draine} \&
  {Weingartner}}{1996}]{Draine:1996ab}
{Draine} B.~T.,  {Weingartner} J.~C.,  1996, \mn@doi [\apj] {10.1086/177887},
  \href {https://ui.adsabs.harvard.edu/abs/1996ApJ...470..551D} {470, 551}

\bibitem[\protect\citeauthoryear{{Ferrara}, {Hirashita}, {Ouchi}  \&
  {Fujimoto}}{{Ferrara} et~al.}{2017}]{Ferrara:2017aa}
{Ferrara} A.,  {Hirashita} H.,  {Ouchi} M.,   {Fujimoto} S.,  2017, \mn@doi
  [\mnras] {10.1093/mnras/stx1898}, \href
  {http://adsabs.harvard.edu/abs/2017MNRAS.471.5018F} {471, 5018}

\bibitem[\protect\citeauthoryear{{Goto} et~al.,}{{Goto}
  et~al.}{2010}]{Goto:2010aa}
{Goto} T.,  et~al., 2010, \mn@doi [\aap] {10.1051/0004-6361/200913182}, \href
  {https://ui.adsabs.harvard.edu/abs/2010A&A...514A...6G} {514, A6}

\bibitem[\protect\citeauthoryear{{Gould} \& {Salpeter}}{{Gould} \&
  {Salpeter}}{1963}]{Gould:1963aa}
{Gould} R.~J.,  {Salpeter} E.~E.,  1963, \mn@doi [\apj] {10.1086/147654}, \href
  {http://adsabs.harvard.edu/abs/1963ApJ...138..393G} {138, 393}

\bibitem[\protect\citeauthoryear{{Guillet}, {Pineau Des For{\^e}ts}  \&
  {Jones}}{{Guillet} et~al.}{2011}]{Guillet:2011aa}
{Guillet} V.,  {Pineau Des For{\^e}ts} G.,   {Jones} A.~P.,  2011, \mn@doi
  [\aap] {10.1051/0004-6361/201015973}, \href
  {http://adsabs.harvard.edu/abs/2011A%26A...527A.123G} {527, A123}

\bibitem[\protect\citeauthoryear{{Hirashita} \& {Aoyama}}{{Hirashita} \&
  {Aoyama}}{2019}]{Hirashita:2019aa}
{Hirashita} H.,  {Aoyama} S.,  2019, \mn@doi [\mnras] {10.1093/mnras/sty2838},
  \href {http://adsabs.harvard.edu/abs/2019MNRAS.482.2555H} {482, 2555}

\bibitem[\protect\citeauthoryear{{Hirashita} \& {Harada}}{{Hirashita} \&
  {Harada}}{2017}]{Hirashita:2017aa}
{Hirashita} H.,  {Harada} N.,  2017, \mn@doi [\mnras] {10.1093/mnras/stx118},
  \href {https://ui.adsabs.harvard.edu/abs/2017MNRAS.467..699H} {467, 699}

\bibitem[\protect\citeauthoryear{{Hirashita} \& {Ichikawa}}{{Hirashita} \&
  {Ichikawa}}{2009}]{Hirashita:2009ac}
{Hirashita} H.,  {Ichikawa} T.~T.,  2009, \mn@doi [\mnras]
  {10.1111/j.1365-2966.2009.14726.x}, \href
  {http://adsabs.harvard.edu/abs/2009MNRAS.396..500H} {396, 500}

\bibitem[\protect\citeauthoryear{{Hirashita} \& {Kobayashi}}{{Hirashita} \&
  {Kobayashi}}{2013}]{Hirashita:2013ab}
{Hirashita} H.,  {Kobayashi} H.,  2013, \mn@doi [Earth, Planets, and Space]
  {10.5047/eps.2013.03.008}, \href
  {http://adsabs.harvard.edu/abs/2013EP%26S...65.1083H} {65, 1083}

\bibitem[\protect\citeauthoryear{{Hirashita} \& {Murga}}{{Hirashita} \&
  {Murga}}{2020}]{Hirashita:2020aa}
{Hirashita} H.,  {Murga} M.~S.,  2020, \mn@doi [\mnras]
  {10.1093/mnras/stz3640}, \href
  {https://ui.adsabs.harvard.edu/abs/2020MNRAS.492.3779H} {492, 3779}

\bibitem[\protect\citeauthoryear{{Hirashita}, {Burgarella}  \&
  {Bouwens}}{{Hirashita} et~al.}{2017}]{Hirashita:2017ab}
{Hirashita} H.,  {Burgarella} D.,   {Bouwens} R.~J.,  2017, \mn@doi [\mnras]
  {10.1093/mnras/stx2349}, \href
  {http://adsabs.harvard.edu/abs/2017MNRAS.472.4587H} {472, 4587}

\bibitem[\protect\citeauthoryear{{Hoang}}{{Hoang}}{2019}]{Hoang:2019aa}
{Hoang} T.,  2019, \mn@doi [\apj] {10.3847/1538-4357/ab1075}, \href
  {https://ui.adsabs.harvard.edu/abs/2019ApJ...876...13H} {876, 13}

\bibitem[\protect\citeauthoryear{{Hoang} \& {Lazarian}}{{Hoang} \&
  {Lazarian}}{2008}]{Hoang:2008aa}
{Hoang} T.,  {Lazarian} A.,  2008, \mn@doi [\mnras]
  {10.1111/j.1365-2966.2008.13249.x}, \href
  {https://ui.adsabs.harvard.edu/abs/2008MNRAS.388..117H} {388, 117}

\bibitem[\protect\citeauthoryear{{Hoang} \& {Lazarian}}{{Hoang} \&
  {Lazarian}}{2009}]{Hoang:2009aa}
{Hoang} T.,  {Lazarian} A.,  2009, \mn@doi [\apj]
  {10.1088/0004-637X/695/2/1457}, \href
  {https://ui.adsabs.harvard.edu/abs/2009ApJ...695.1457H} {695, 1457}

\bibitem[\protect\citeauthoryear{{Hoang}, {Tram}, {Lee}  \& {Ahn}}{{Hoang}
  et~al.}{2019}]{Hoang:2019ab}
{Hoang} T.,  {Tram} L.~N.,  {Lee} H.,   {Ahn} S.-H.,  2019, \mn@doi [Nature
  Astronomy] {10.1038/s41550-019-0763-6}, \href
  {https://ui.adsabs.harvard.edu/abs/2019NatAs...3..766H} {3, 766}

\bibitem[\protect\citeauthoryear{{Hou}, {Hirashita}  \& {Micha{\l}owski}}{{Hou}
  et~al.}{2016}]{Hou:2016aa}
{Hou} K.-C.,  {Hirashita} H.,   {Micha{\l}owski} M.~J.,  2016, \mn@doi [\pasj]
  {10.1093/pasj/psw085}, \href
  {http://adsabs.harvard.edu/abs/2016PASJ...68...94H} {68, 94}

\bibitem[\protect\citeauthoryear{{Inoue}}{{Inoue}}{2005}]{Inoue:2005aa}
{Inoue} A.~K.,  2005, \mn@doi [\mnras] {10.1111/j.1365-2966.2005.08890.x},
  \href {http://adsabs.harvard.edu/abs/2005MNRAS.359..171I} {359, 171}

\bibitem[\protect\citeauthoryear{{Jones}, {Tielens}, {Hollenbach}  \&
  {McKee}}{{Jones} et~al.}{1994}]{Jones:1994aa}
{Jones} A.~P.,  {Tielens} A.~G.~G.~M.,  {Hollenbach} D.~J.,   {McKee} C.~F.,
  1994, \mn@doi [\apj] {10.1086/174689}, \href
  {http://adsabs.harvard.edu/abs/1994ApJ...433..797J} {433, 797}

\bibitem[\protect\citeauthoryear{{Jones}, {Tielens}  \& {Hollenbach}}{{Jones}
  et~al.}{1996}]{Jones:1996aa}
{Jones} A.~P.,  {Tielens} A.~G.~G.~M.,   {Hollenbach} D.~J.,  1996, \mn@doi
  [\apj] {10.1086/177823}, \href
  {http://adsabs.harvard.edu/abs/1996ApJ...469..740J} {469, 740}

\bibitem[\protect\citeauthoryear{{Kobayashi} \& {Tanaka}}{{Kobayashi} \&
  {Tanaka}}{2010}]{Kobayashi:2010aa}
{Kobayashi} H.,  {Tanaka} H.,  2010, \mn@doi [\icarus]
  {10.1016/j.icarus.2009.10.004}, \href
  {http://adsabs.harvard.edu/abs/2010Icar..206..735K} {206, 735}

\bibitem[\protect\citeauthoryear{{Larson} \& {Tinsley}}{{Larson} \&
  {Tinsley}}{1978}]{Larson:1978aa}
{Larson} R.~B.,  {Tinsley} B.~M.,  1978, \mn@doi [\apj] {10.1086/155753}, \href
  {https://ui.adsabs.harvard.edu/abs/1978ApJ...219...46L} {219, 46}

\bibitem[\protect\citeauthoryear{{Li} \& {Draine}}{{Li} \&
  {Draine}}{2001}]{Li:2001aa}
{Li} A.,  {Draine} B.~T.,  2001, \mn@doi [\apj] {10.1086/323147}, \href
  {http://adsabs.harvard.edu/abs/2001ApJ...554..778L} {554, 778}

\bibitem[\protect\citeauthoryear{{Liffman} \& {Clayton}}{{Liffman} \&
  {Clayton}}{1989}]{Liffman:1989aa}
{Liffman} K.,  {Clayton} D.~D.,  1989, \mn@doi [\apj] {10.1086/167440}, \href
  {https://ui.adsabs.harvard.edu/abs/1989ApJ...340..853L} {340, 853}

\bibitem[\protect\citeauthoryear{{Lim} et~al.,}{{Lim}
  et~al.}{2020}]{Lim:2020aa}
{Lim} C.-F.,  et~al., 2020, \mn@doi [\apj] {10.3847/1538-4357/ab607f}, \href
  {https://ui.adsabs.harvard.edu/abs/2020ApJ...889...80L} {889, 80}

\bibitem[\protect\citeauthoryear{{Liu} \& {Hirashita}}{{Liu} \&
  {Hirashita}}{2019}]{Liu:2019aa}
{Liu} H.-M.,  {Hirashita} H.,  2019, \mn@doi [\mnras] {10.1093/mnras/stz2647},
  \href {https://ui.adsabs.harvard.edu/abs/2019MNRAS.490..540L} {490, 540}

\bibitem[\protect\citeauthoryear{{Mathis}, {Rumpl}  \& {Nordsieck}}{{Mathis}
  et~al.}{1977}]{Mathis:1977aa}
{Mathis} J.~S.,  {Rumpl} W.,   {Nordsieck} K.~H.,  1977, \mn@doi [\apj]
  {10.1086/155591}, \href {http://adsabs.harvard.edu/abs/1977ApJ...217..425M}
  {217, 425}

\bibitem[\protect\citeauthoryear{{Mathis}, {Mezger}  \& {Panagia}}{{Mathis}
  et~al.}{1983}]{Mathis:1983aa}
{Mathis} J.~S.,  {Mezger} P.~G.,   {Panagia} N.,  1983, \aap, \href
  {http://adsabs.harvard.edu/abs/1983A%26A...128..212M} {128, 212}

\bibitem[\protect\citeauthoryear{{McKinnon}, {Vogelsberger}, {Torrey},
  {Marinacci}  \& {Kannan}}{{McKinnon} et~al.}{2018}]{McKinnon:2018aa}
{McKinnon} R.,  {Vogelsberger} M.,  {Torrey} P.,  {Marinacci} F.,   {Kannan}
  R.,  2018, \mn@doi [\mnras] {10.1093/mnras/sty1248}, \href
  {http://adsabs.harvard.edu/abs/2018MNRAS.478.2851M} {478, 2851}

\bibitem[\protect\citeauthoryear{{McKinnon}, {Kannan}, {Vogelsberger},
  {O'Neil}, {Torrey}  \& {Li}}{{McKinnon} et~al.}{2020}]{McKinnon:2020aa}
{McKinnon} R.,  {Kannan} R.,  {Vogelsberger} M.,  {O'Neil} S.,  {Torrey} P.,
  {Li} H.,  2020, \mnras, \href
  {https://ui.adsabs.harvard.edu/abs/2019arXiv191202825M} {submitted,
  arXiv:1912.02825}

\bibitem[\protect\citeauthoryear{{Narayanan}, {Conroy}, {Dav{\'e}}, {Johnson}
  \& {Popping}}{{Narayanan} et~al.}{2018}]{Narayanan:2018aa}
{Narayanan} D.,  {Conroy} C.,  {Dav{\'e}} R.,  {Johnson} B.~D.,   {Popping} G.,
   2018, \mn@doi [\apj] {10.3847/1538-4357/aaed25}, \href
  {http://adsabs.harvard.edu/abs/2018ApJ...869...70N} {869, 70}

\bibitem[\protect\citeauthoryear{{Nozawa} \& {Fukugita}}{{Nozawa} \&
  {Fukugita}}{2013}]{Nozawa:2013aa}
{Nozawa} T.,  {Fukugita} M.,  2013, \mn@doi [\apj]
  {10.1088/0004-637X/770/1/27}, \href
  {http://adsabs.harvard.edu/abs/2013ApJ...770...27N} {770, 27}

\bibitem[\protect\citeauthoryear{{Nozawa}, {Asano}, {Hirashita}  \&
  {Takeuchi}}{{Nozawa} et~al.}{2015}]{Nozawa:2015aa}
{Nozawa} T.,  {Asano} R.~S.,  {Hirashita} H.,   {Takeuchi} T.~T.,  2015,
  \mn@doi [\mnras] {10.1093/mnrasl/slu175}, \href
  {http://adsabs.harvard.edu/abs/2015MNRAS.447L..16N} {447, L16}

\bibitem[\protect\citeauthoryear{{O'Donnell} \& {Mathis}}{{O'Donnell} \&
  {Mathis}}{1997}]{ODonnell:1997aa}
{O'Donnell} J.~E.,  {Mathis} J.~S.,  1997, \mn@doi [\apj] {10.1086/303903},
  \href {https://ui.adsabs.harvard.edu/abs/1997ApJ...479..806O} {479, 806}

\bibitem[\protect\citeauthoryear{{Ormel}, {Paszun}, {Dominik}  \&
  {Tielens}}{{Ormel} et~al.}{2009}]{Ormel:2009aa}
{Ormel} C.~W.,  {Paszun} D.,  {Dominik} C.,   {Tielens} A.~G.~G.~M.,  2009,
  \mn@doi [\aap] {10.1051/0004-6361/200811158}, \href
  {http://adsabs.harvard.edu/abs/2009A%26A...502..845O} {502, 845}

\bibitem[\protect\citeauthoryear{{Pei}}{{Pei}}{1992}]{Pei:1992aa}
{Pei} Y.~C.,  1992, \mn@doi [\apj] {10.1086/171637}, \href
  {http://adsabs.harvard.edu/abs/1992ApJ...395..130P} {395, 130}

\bibitem[\protect\citeauthoryear{{Rau}, {Hirashita}  \& {Murga}}{{Rau}
  et~al.}{2019}]{Rau:2019aa}
{Rau} S.-J.,  {Hirashita} H.,   {Murga} M.,  2019, \mn@doi [\mnras]
  {10.1093/mnras/stz2532}, \href
  {https://ui.adsabs.harvard.edu/abs/2019MNRAS.489.5218R} {489, 5218}

\bibitem[\protect\citeauthoryear{{Sanders} \& {Mirabel}}{{Sanders} \&
  {Mirabel}}{1996}]{Sanders:1996aa}
{Sanders} D.~B.,  {Mirabel} I.~F.,  1996, \mn@doi [\araa]
  {10.1146/annurev.astro.34.1.749}, \href
  {https://ui.adsabs.harvard.edu/abs/1996ARA&A..34..749S} {34, 749}

\bibitem[\protect\citeauthoryear{{Schreiber}, {Elbaz}, {Pannella}, {Ciesla},
  {Wang}  \& {Franco}}{{Schreiber} et~al.}{2018}]{Schreiber:2018aa}
{Schreiber} C.,  {Elbaz} D.,  {Pannella} M.,  {Ciesla} L.,  {Wang} T.,
  {Franco} M.,  2018, \mn@doi [\aap] {10.1051/0004-6361/201731506}, \href
  {https://ui.adsabs.harvard.edu/abs/2018A&A...609A..30S} {609, A30}

\bibitem[\protect\citeauthoryear{{Seok}, {Hirashita}  \& {Asano}}{{Seok}
  et~al.}{2014}]{Seok:2014aa}
{Seok} J.~Y.,  {Hirashita} H.,   {Asano} R.~S.,  2014, \mn@doi [\mnras]
  {10.1093/mnras/stu120}, \href
  {http://adsabs.harvard.edu/abs/2014MNRAS.439.2186S} {439, 2186}

\bibitem[\protect\citeauthoryear{{Seon} \& {Draine}}{{Seon} \&
  {Draine}}{2016}]{Seon:2016aa}
{Seon} K.-I.,  {Draine} B.~T.,  2016, \mn@doi [\apj]
  {10.3847/1538-4357/833/2/201}, \href
  {https://ui.adsabs.harvard.edu/abs/2016ApJ...833..201S} {833, 201}

\bibitem[\protect\citeauthoryear{{Symeonidis} et~al.,}{{Symeonidis}
  et~al.}{2013}]{Symeonidis:2013aa}
{Symeonidis} M.,  et~al., 2013, \mn@doi [\mnras] {10.1093/mnras/stt330}, \href
  {https://ui.adsabs.harvard.edu/abs/2013MNRAS.431.2317S} {431, 2317}

\bibitem[\protect\citeauthoryear{{Takeuchi}, {Ishii}, {Nozawa}, {Kozasa}  \&
  {Hirashita}}{{Takeuchi} et~al.}{2005a}]{Takeuchi:2005aa}
{Takeuchi} T.~T.,  {Ishii} T.~T.,  {Nozawa} T.,  {Kozasa} T.,   {Hirashita} H.,
   2005a, \mn@doi [\mnras] {10.1111/j.1365-2966.2005.09337.x}, \href
  {http://adsabs.harvard.edu/abs/2005MNRAS.362..592T} {362, 592}

\bibitem[\protect\citeauthoryear{{Takeuchi}, {Buat}  \&
  {Burgarella}}{{Takeuchi} et~al.}{2005b}]{Takeuchi:2005ab}
{Takeuchi} T.~T.,  {Buat} V.,   {Burgarella} D.,  2005b, \mn@doi [\aap]
  {10.1051/0004-6361:200500158}, \href
  {https://ui.adsabs.harvard.edu/abs/2005A&A...440L..17T} {440, L17}

\bibitem[\protect\citeauthoryear{{Tanaka}, {Inaba}  \& {Nakazawa}}{{Tanaka}
  et~al.}{1996}]{Tanaka:1996aa}
{Tanaka} H.,  {Inaba} S.,   {Nakazawa} K.,  1996, \mn@doi [\icarus]
  {10.1006/icar.1996.0170}, \href
  {https://ui.adsabs.harvard.edu/abs/1996Icar..123..450T} {123, 450}

\bibitem[\protect\citeauthoryear{{Tielens}, {McKee}, {Seab}  \&
  {Hollenbach}}{{Tielens} et~al.}{1994}]{Tielens:1994aa}
{Tielens} A.~G.~G.~M.,  {McKee} C.~F.,  {Seab} C.~G.,   {Hollenbach} D.~J.,
  1994, \mn@doi [\apj] {10.1086/174488}, \href
  {http://adsabs.harvard.edu/abs/1994ApJ...431..321T} {431, 321}

\bibitem[\protect\citeauthoryear{{Toba} et~al.,}{{Toba}
  et~al.}{2020}]{Toba:2020aa}
{Toba} Y.,  et~al., 2020, \mn@doi [\apj] {10.3847/1538-4357/ab616d}, \href
  {https://ui.adsabs.harvard.edu/abs/2020ApJ...889...76T} {889, 76}

\bibitem[\protect\citeauthoryear{{Weingartner} \& {Draine}}{{Weingartner} \&
  {Draine}}{2001}]{Weingartner:2001aa}
{Weingartner} J.~C.,  {Draine} B.~T.,  2001, \mn@doi [\apj] {10.1086/318651},
  \href {http://adsabs.harvard.edu/abs/2001ApJ...548..296W} {548, 296}

\bibitem[\protect\citeauthoryear{{Williams} \& {Wetherill}}{{Williams} \&
  {Wetherill}}{1994}]{Williams:1994aa}
{Williams} D.~R.,  {Wetherill} G.~W.,  1994, \mn@doi [\icarus]
  {10.1006/icar.1994.1010}, \href
  {https://ui.adsabs.harvard.edu/abs/1994Icar..107..117W} {107, 117}

\bibitem[\protect\citeauthoryear{{Witt} \& {Gordon}}{{Witt} \&
  {Gordon}}{2000}]{Witt:2000aa}
{Witt} A.~N.,  {Gordon} K.~D.,  2000, \mn@doi [\apj] {10.1086/308197}, \href
  {https://ui.adsabs.harvard.edu/abs/2000ApJ...528..799W} {528, 799}

\bibitem[\protect\citeauthoryear{{Yamasawa}, {Habe}, {Kozasa}, {Nozawa},
  {Hirashita}, {Umeda}  \& {Nomoto}}{{Yamasawa} et~al.}{2011}]{Yamasawa:2011aa}
{Yamasawa} D.,  {Habe} A.,  {Kozasa} T.,  {Nozawa} T.,  {Hirashita} H.,
  {Umeda} H.,   {Nomoto} K.,  2011, \mn@doi [\apj]
  {10.1088/0004-637X/735/1/44}, \href
  {http://adsabs.harvard.edu/abs/2011ApJ...735...44Y} {735, 44}

\bibitem[\protect\citeauthoryear{{Zavala} et~al.,}{{Zavala}
  et~al.}{2018}]{Zavala:2018aa}
{Zavala} J.~A.,  et~al., 2018, \mn@doi [\mnras] {10.1093/mnras/sty217}, \href
  {https://ui.adsabs.harvard.edu/abs/2018MNRAS.475.5585Z} {475, 5585}

\bibitem[\protect\citeauthoryear{{Zubko}, {Mennella}, {Colangeli}  \&
  {Bussoletti}}{{Zubko} et~al.}{1996}]{Zubko:1996aa}
{Zubko} V.~G.,  {Mennella} V.,  {Colangeli} L.,   {Bussoletti} E.,  1996,
  \mn@doi [\mnras] {10.1093/mnras/282.4.1321}, \href
  {http://adsabs.harvard.edu/abs/1996MNRAS.282.1321Z} {282, 1321}

\makeatother
\end{thebibliography}


\appendix

\section{Analogy between shattering and rotational disruption}\label{app:analog}

Since both shattering and rotational disruption produce grain fragments, the equations
describing them should have some similarity. Here, we make an attempt of deriving
equation (\ref{eq:dis}) from the shattering equation.
Shattering is described by the following equation in our model (HM20;
developed from \citealt{Jones:1994aa,Jones:1996aa}):
\begin{align}
\left[\frac{\partial\rho_\mathrm{d}(m,\, t)}{\partial t}\right]_\mathrm{shat}
= -m\rho_\mathrm{d}(m,\, t)\int_0^\infty\alpha (m_1,\, m)\rho_\mathrm{d}(m_1,\, t)
\mathrm{d}m_1\nonumber\\
+ \int_0^\infty\int_0^\infty\alpha (m_1,\, m_2)\rho_\mathrm{d}(m_1,\, t)\rho_\mathrm{d}(m_2,\, t)
\mu_\mathrm{shat}(m;\, m_1,\, m_2)\mathrm{d}m_1
\mathrm{d}m_2,\label{eq:shat}
\end{align}
where $\mu_\mathrm{shat}$ describes the grain mass distribution function of
the shattered fragments produced from a grain with mass $m_1$ in
the collision with a grain with mass $m_2$, and
$\alpha$ is written as
\begin{align}
\alpha (m_1,\, m_2)\equiv\frac{\sigma_{1,2}v_{1,2}}{m_1m_2},\label{eq:alpha}
\end{align}
where $\sigma_{1,2}$ and $v_{1,2}$ are the collisional cross-section
and the relative velocity between the
two colliding grains (with masses $m_1$ and $m_2$), respectively.
We expect that, if we replace the grain--grain
collision time-scale with the rotational disruption time-scale, we obtain
the equation that describes rotational disruption. A grain with mass $m$
experiences collisions on a time-scale of $\tau_\mathrm{coll}$, which is
evaluated as
\begin{align}
\tau_\mathrm{coll}(m)^{-1}=m\int_0^\infty\alpha (m',\, m)\rho_\mathrm{d}(m',\, t)\,\mathrm{d}m'.
\end{align}
In rotational disruption, the grain $m_1$ spontaneously fragments; thus,
the fragment mass distribution function does not depend on $m_2$.
Eventually, if we convert $\tau_\mathrm{coll}$ to $\tau_\mathrm{disr}$
and $\mu_\mathrm{shat}(m;\, m_1,\, m_2)$ to $\theta_\mathrm{disr}(m;\, m_1)$, we obtain the
evolution of grain size distribution by rotational disruption (equation \ref{eq:dis}).

\bsp	
\label{lastpage}
\end{document}